\documentclass[]{interact}

\usepackage{xcolor}
\usepackage{url}
\usepackage{epstopdf}% To incorporate .eps illustrations using PDFLaTeX, etc.
\usepackage{appendix}
\usepackage{subfigure}% Support for small, `sub' figures and tables

\usepackage{float}
\usepackage{natbib}% Citation support using natbib.sty
\bibpunct[, ]{(}{)}{;}{a}{}{,}% Citation support using natbib.sty
\renewcommand\bibfont{\fontsize{10}{12}\selectfont}% Bibliography support using natbib.sty

\theoremstyle{plain}% Theorem-like structures provided by amsthm.sty

\theoremstyle{definition}

\theoremstyle{remark}

\newcommand{\bu}{{\boldsymbol{u}}}
\newcommand{\grad}{\nabla}
\newcommand{\bcdot}{\boldsymbol{\cdot}}
\renewcommand{\div}{\grad\bcdot}  % Overwrites \div = divide-by symbol

\newcommand{\lap}{\nabla^2}

\providecommand\pa{{\partial}}

\begin{document}

\title{Radiatively-Cooled Moist Convection under an Idealised Climate Change Scenario: Linear Analysis}

\author{G. N. Dritschel$^{1}$$^{\ast}$\thanks{$^\ast$Corresponding author. Email: mmgnd@leeds.ac.uk
\vspace{6pt}}, S. M. Tobias$^{1}$, D. J. Parker$^{1,2,3}$ and L. Tomassini$^{4}$\\\vspace{6pt}  $^{1}$School of Mathematics, University of Leeds, Leeds LS2 9JT, UK\\ $^{2}$NORCE Norwegian Research Centre AS, Bergen NO-5838, NO\\ $^{3}$National Centre for Atmospheric Science (NCAS), Leeds LS2 9JT\\$^{4}$Met Office, Exeter EX1 3PB, UK \\\vspace{6pt}\received{v4.4 released October 2012} }

\maketitle

\begin{abstract}
In order to explore the effects of climate change on atmospheric convection and the water cycle, we develop and analyse an extension of the Rainy-B\'enard model, which is itself a moist version of the Rayleigh-B\'enard model of dry convection. Including moisture changes the character of the convection, with condensation providing a source of buoyancy via latent heating. The climate change model is set up by imposing a variable radiative cooling rate, prescribing surface temperature and relative humidity, and imposing a moist-pseudoadiabatic profile at the top boundary (a flux boundary condition). The model is analysed across the climate parameter space by examining diagnostics of the model's basic state, and its stability, with Convective Available Potential Energy (CAPE) calculations and a linear stability analysis. We use the linear stability results to identify new parameters relevant for this moist convective system, and to understand how the linear instability responds to the climate parameters. In particular, we define the "Rainy number" as a scaled ratio of positive-area CAPE and diffusion parameters. An alternative radiative-based Rainy number also is shown to describe the parameter space, especially for problems relating to changes in flux conditions. The analysis provides a novel theoretical understanding of how the dynamics and scales of moist convection and hence precipitation will change under climate change.  \\
\end{abstract}


\section{Introduction}
The changes in atmospheric deep convection, and the character of its associated rainfall, are uncertain in future climate \citep{OGormann_PE}. Moist convection is a key source of uncertainty in climate model simulations, due to its small scale processes, and its feedback on large scale climate processes, such as the Hadley circulation \citep{Tomassini_2022}. Understanding changes in the intensity, structure and organisation of moist convection under climate change are key for improving climate model projections, which are pertinent tools for protecting vulnerable communities in the future. How moist convection changes under climate change is therefore a critically-important problem in climate research.
\\
\newline
Changes in deep convection have been investigated using a range of approaches. One approach involves the use of ensemble data from global-scale General Circulation Models (GCMs), with energy and moisture budgets used to examine future rainfall change \citep{HeldSoden_GCM,ByrneOGormann_GCM,Seager_GCM}. However there is a large uncertainty in modelled convective rainfall over the tropics, due to coarse resolutions ($\sim$ 50km) which require subgrid-scale convective parametrisations \citep{OGormann_PE}. With the development of exascale computing, there have been significant advances in GCM resolution, with kilometre-scale (k-scale) models now being able to explicitly resolve convection on global domains \citep{Schar_Kscale}. Recently such models have been used in the study of rainfall change \citep{Cheng_Kscale}. However moist biases have been found in studies of global convection-permitting models (CPMs), potentially due to these models not capturing key convective processes such as turbulence \citep{TomassiniGrayZone}. Initial k-scale studies indicate that there still remain gaps in our understanding of atmospheric moist convection, some of which can be investigated using idealised studies which can resolve turbulence \citep{Guichard_Rev}.
\\
\newline
An alternative to global-scale modelling involves using CRMs (Cloud Resolving Models) or LES (Large Eddy Simulations) to resolve explicitly convective clouds, and their associated transient motions on smaller domains (e.g. 100km$\times$20km). Both GCMs and CRMs are dynamically based on an approximation of the Navier-Stokes equations, called the ``primitive equations" \citep{Phillips_1966}, CRMs typically have higher resolutions (10m-1km) than GCMs. Although CRMs can marginally resolve convection, complexity arises in these models by the choice of the parametrisations used for the subgrid-scale (i) turbulent motions, (ii) microphysical processes and (iii) radiative processes \citep{Guichard_Rev}. A limitation in CRMs is the need to prescribe lateral boundary conditions, which are not required in GCMs. CRMs can be used for idealised studies of convection, including the development of diagnostics (e.g.\ \cite{MullerOGormannBack_CRM}, \cite{MullerTaka_CRM}). The study by \cite{Bretherton_CRM} used LES to test climate change sensitivities (e.g.\ to surface temperature warming). CRMs and LES have been used to inform parametrisations of subgrid-scale processes in more complex models \citep{deRoode2016large,guichard2004modelling}. Another use of CRMs lies in the accessibility of 4D (3D + time)
fields of moisture, vertical wind, etc.\ which can be used to complement (limited) data from observations \citep{oue2016estimation}. 
\\
\newline
In the present paper, we present a simple turbulence-resolving model, stemming from an idealised model of moist convection (the Rainy-B\'enard model, from \cite{vallis2019simple}). The simplicity of the model, the radiation term and uniform boundary conditions therein allow us to establish a theoretical framework required for a conceptual approach to the climate change problem, based on fundamental principles of fluid dynamics. In particular, we will link a wider range of moist, radiatively cooled regimes, to understand the underlying mathematical behaviour of our climate change scenario.
\\
\newline
We aim first to document a well-posed climate scenario with Rainy-B\'enard convection, through the choice of a radiative cooling parameterisation and boundary conditions. Second, we identify key parameters for the climate problem, particularly a Rayleigh number that is dependent on moisture and radiation. Third, we document the model's equilibrium  (basic state) solutions and their stability, by calculation of both the basic state's convective available potential energy and its linear stability. Finally, we discuss the linear response of the model to radiative and thermal climate change. 
\\
\newline
Convection is in general characterised by highly non-linear behaviour. However, linear theory has been used to understand the non-linear regimes of classical dry Rayleigh-Be\'nard convection \citep{christopher_lebars_llewellynsmith_2023}, and therefore, we perform a linear analysis of the (moist) Rainy-B\'enard model to gain an initial insight of the behaviour underpinning the non-linear regimes of the model. The study by \cite{Agasthya_2024} used non-linear simulations of radiatively cooled Rainy-Benard convection, found similar scalings to those found in previous CRM studies, indicating that the Rainy-B\'enard model can provide relevant results despite its simplicity. A linear analysis of the Rainy-Be\'nard model has been conducted in the \cite{oishi2024_lsa} study, which examined the convective onset (critical Rayleigh number), convective instability and moist internal gravity waves for the Rainy-Be\'nard model. We present a similar linear analysis here, however we use alternative boundary conditions and include a radiative cooling term as required for a climate change study. We also conduct a more detailed analytical treatment of the basic state solution, use a different approach to solving the linear stability eigenvalue problem, and provide a different dispersion relationship for the (linear) internal gravity waves, and use conditional instability and moisture diagnostics to interpret the results. The linear results presented in this paper provides the basis for a future investigation into the non-linear behaviour of moist convection under climate change, as has been successfully demonstrated for the ``dry'' classical Rayleigh-Be\'nard problem over many years \citep{1961Chandrasekhar,christopher_lebars_llewellynsmith_2023}.
\\
\newline
The paper starts with setting up the Rainy-B\'enard model for a climate change study, discussing boundary conditions, radiative cooling, different non-dimensionalisations and key parameters (Section 2). The analytical basic state solution is calculated across the climate parameter space, then some key diagnostics are presented (Section 3). We then analyse the stability of the basic state and its dependence on the climate parameters, using (i) CAPE calculations and (ii) a linear stability analysis (Section 4). The main findings and future research involving non-linear simulations are discussed in the conclusion (Section 5).

\section{Model Set Up for Climate-Forcing Simulations}
We use a radiatively extended Rainy-Benard model for this study. The model equations are the same as those in \cite{Agasthya_2024}, but different boundary conditions are applied. The model set up differs from \cite{oishi2024_lsa} both in the inclusion of radiative cooling, and in the boundary conditions. The Boussinesq, radiative moist convective system is given by,
\begin{align}
    \frac{D\mathbf{{u}}}{D{t}} &= - \mathbf{{\nabla}} {\phi} + {b}\mathbf{k} + \nu \mathbf{{\nabla}^2{u}},
    \label{momentum} \\
    \frac{D{b}}{D{t}} &= {\gamma}\frac{{q}-{q_s}}{{\tau}}\mathcal{H}({q}-{q_s})+\kappa\mathbf{{\nabla}^2}{b}-\frac{g}{\theta_0}{r},
    \label{buoyancy} \\
    \frac{D{q}}{D{t}} &= -\frac{{q}-{q_s}}{{\tau}}\mathcal{H}({q}-{q_s})+\kappa_q\mathbf{{\nabla}^2}{q}, 
    \label{specific humidity} \\
    \delta T &= \frac{\theta_0}{g}b - \frac{g}{c_p}z, \label{temp}\\
    {q_s} &= q_0 e^{{\alpha}\delta T}, \label{sat specific humidity} \\
    \mathbf{{\nabla} \cdot {u}} &= 0. \label{incomp}
\end{align}
The model has idealised microphysics, with saturated water vapor ($q > q_s$) being removed at the condensational timescale $\tau$. There is no liquid water or ice phase included in this formulation. Radiation is also idealised as a constant cooling rate, $r$, which provides a first-order approximation of the radiative cooling in the atmosphere \citep{SimpleSpectralModelsforAtmosphericRadiativeCooling, Agasthya_2024}. The buoyancy equation differs from dry Rayleigh-B\'enard convection in the inclusion of a condensation term, $C \equiv \gamma({{q}-{q_s}})\mathcal{H}({q}-{q_s})/\tau$ \citep{vallis2019simple}, and has been extended here for climate study with a bulk radiative cooling term, $gr/\theta_0$. Note condensation also provides a sink in moisture (equation \eqref{specific humidity}). The relationship of the perturbation temperature ($\delta T = T-T_0$, where $T_0 = 300K$) to the buoyancy is given by equation \eqref{temp}, and can be derived from the first law of thermodynamics \citep{vallis2019simple}. Equation \eqref{sat specific humidity} is the Clausius-Clapeyron relation \eqref{sat specific humidity}, linearised about $T_0$. The simplicity of the radiative cooling and microphysics makes a detailed fundamental analysis of the model possible.
\\
\newline

\subsection{Idealised Climate Change - Boundary Conditions}
\label{BC section}
An idealised climate change scenario can be set up by careful choice of the boundary conditions. We take the surface boundary conditions to be
\begin{equation}
    b(0) = g\Delta T_{\textsf{surf}}/\theta_0, \quad q(0) = RH_{\textsf{surf}} \:q_s(0)=RH_{\textsf{surf}}\:q_0e^{\alpha\Delta T_{\textsf{surf}}}, \quad \mathbf{u}(0) = \mathbf{0}.
    \label{surface BCs}
\end{equation}
We impose a surface temperature of $T(0) = T_0 + \Delta T_{\textsf{surf}}$, and a surface relative humidity, where the relative humidity is defined as $RH \equiv q/q_s$. We use equation \eqref{temp} to define $b(0)$, and equation \eqref{sat specific humidity} to define $q(0)$. We also use idealised no-slip boundary conditions, at both boundaries (as in \cite{vallis2019simple}). 
\\
\newline
The top of the domain represents the tropopause. We use moist pseudoadiabatic boundary conditions at the top boundary, which can be expressed as,
\begin{equation}
    \frac{dm}{dz}(H) = \frac{db}{dz}(H) + \gamma \frac{dq}{dz}(H) = 0, \quad q(H) = q_s(H), \quad \mathbf{u}(H) = \mathbf{0}.
    \label{top BCs}
\end{equation}
Note that $m = b + \gamma q$ is the moist static energy of the model, and the moist pseudoadiabat is defined as a profile for which $dm/dz = 0$ and $q = q_s$ \citep{vallis2019simple}. Imposing moist pseudoadiabatic boundary conditions allows the temperature and moisture to adjust to moist convection. In \cite{vallis2019simple}, \cite{oishi2024_lsa} and \cite{Agasthya_2024}, both $T(H)$ and $q(H)$ were fixed, so there was no possibility to adjust the upper levels to a convectively controlled profile, and the stratification leads to an unrealistic downward heat flux. In the real atmosphere, the diffusive fluxes of buoyancy and humidity ($\kappa \, \partial b/\partial z, \: \kappa_q \, \partial q/\partial z$) are small at the tropopause, and the profile is close to the moist pseudoadiabat, especially in the tropics. The top boundary conditions are consistent with atmospheric conditions at the tropopause, and they additionally allow convective adjustment of the profiles.
\\
\newline
The idealised climate change scenario is imposed by increasing both radiative cooling and surface temperature, inline with Figure 4 of \cite{SimpleSpectralModelsforAtmosphericRadiativeCooling}. We consider climate change (increasing both $r \:\&\: \Delta T_{\textsf{surf}}$) for a range of different surface relative humidities in the following analysis.

\subsection{Dry Adiabatic Buoyancy Non-Dimensionalisation}
Following Appendix 7.1 in \cite{vallis2019simple}, we use a ``buoyancy based non-dimensionalisation''. We cannot use the classical Rayleigh-B\'enard temperature scale $[T] = T(0)-T(H)$, since $T(H)$ is free to adjust to moist convection under our moist pseudoadiabtic boundary conditions specified in Section \ref{BC section}. The temperature scale is instead set by the dry adiabatic temperature difference,\ $[T] = T_d(0)-T_d(H) = gH/c_p \sim 100K$.
\\
\newline
We use equation \eqref{temp} to set the buoyancy scale to be proportional to the temperature scale, $[B] = g[T]/\theta_0 \sim 3.3 ms^{-2}$. The length scale is defined to be height of the domain, $[L] = H \sim 10km$, and we define the timescale using $[t] = ([L]/[b])^{1/2} \sim 55s$. We scale the specific humidity by $[q] =q_0 \sim 3.8\times 10^{-3}$ which is lower than its real value (see Appendix \ref{corected parameters}), to ensure that the fluxes of buoyancy and humidity are small at the top boundary. The velocity scale is set by $[U] = [L]/[t] \sim 180 ms^{-1}$ and the pressure scale by $[\phi] = [U]^2 \sim 3.3 \times 10^5 Pa$. The non-dimensional system is then given by,
\begin{align}
    \frac{D\mathbf{{\hat{u}}}}{D{\hat{t}}} &= - \mathbf{{\hat{\nabla}}} \hat{\phi} + \hat{b}\mathbf{k} + \Bigg(\frac{Pr}{Ra}\Bigg)^{\frac{1}{2}}\mathbf{\hat{\nabla}^2\hat{u}},
    \label{dif momentum} \\
    \frac{D\hat{b}}{D\hat{t}} &= \hat{\gamma}\frac{\hat{q}-\hat{q_s}}{\hat{\tau}}\mathcal{H}(\hat{q}-\hat{q_s})+\frac{1}{(PrRa)^{\frac{1}{2}}}\mathbf{\hat{\nabla}^2}\hat{b}-\hat{r},
    \label{dif non-d buoyancy} \\
    \frac{D\hat{q}}{D\hat{t}} &= -\frac{\hat{q}-\hat{q_s}}{\hat{\tau}}\mathcal{H}(\hat{q}-\hat{q_s})+\frac{S_m}{(PrRa)^{\frac{1}{2}}}\mathbf{\hat{\nabla}^2}\hat{q}, 
    \label{dif non-d specific humidity} \\
    \mathbf{\hat{\nabla} \cdot \hat{u}} &= 0, \\
    \hat{\delta T} &= \hat{b} - \hat{z}, 
    \label{T eqn} \\
    \hat{q_s} &= e^{\hat{\alpha}\hat{\delta T}}. \label{dif non-d sat specific humidity}
\end{align}
Here we have the non-dimensional parameters
\begin{equation}
    Pr = \frac{\nu}{\kappa}, \quad Ra = \frac{g^2 H^4}{\theta_0 c_p \nu \kappa}, \quad \hat{\gamma} = \frac{\gamma q_0 \theta_0 c_p}{g^2 H}, \quad \hat{\tau} = \frac{\tau g}{(c_p \theta_0)^{1/2}}
\end{equation}
\begin{equation}
    \hat{r} = \frac{(c_p^3 \theta_0)^{1/2} r}{g^2 H}, \quad S_m = \frac{\kappa_q}{\kappa}, \quad \hat{\alpha} = \frac{gH\alpha}{c_p}, \quad \hat{b}_{\textsf{surf}} = \frac{c_p \Delta T_{\textsf{surf}}}{g H}.
\end{equation}
We take $Pr = S_m = 1$, $\hat{\gamma} = 0.25$, $\hat{\alpha} = 6.0$ and $\hat{\tau} = 0.05$ (corresponding to a condensational timescale of $\tau \sim 1s$). We vary the parameters $\hat{r}$, $\hat{b}_{\textsf{surf}}$ and $Ra$. Note that $\hat{r} = 1\times10^{-5}$ for a typical atmospheric cooling rate of $2K\text{day}^{-1}$, and $\hat{b}_{\textsf{surf}} = 0.05$ for a surface temperature increase of $5 K$. For molecular values of $\kappa$ and $\nu$, $Ra \sim 10^{23}$.
\\
\newline
The dry adiabatic temperature and buoyancy scales used in this non-dimensionalisation do not reflect changes in stability caused by changes in the temperature profile via changes in surface temperature, surface moisture, or ultimately, radiative cooling. Therefore $Ra$ in this non-dimensionalisation is only sensitive to changes in $\kappa$ or $\nu$. However, the dry adiabatic buoyancy non-dimensionalisation simplifies the form of the equations, and allows easy implementation of the climate change scenario, and we therefore employ the dry adiabatic buoyancy non-dimensionalisation in the rest of the analysis in this paper noting that we drop the hats on the parameters for the rest of the analysis.

\subsection{Integral Constraints: Energy and Moisture Budgets}
\label{moiststab}
By integrating the moisture and buoyancy equations over the atmospheric column, we can obtain integral constraints on the model. This approach is well understood in climate science and is one basis for analysing future rainfal patterns \citep{HeldSoden_GCM,ByrneOGormann_GCM}. We consider these constraints for equilibrium, in which $\pa /\pa t = 0$. Using equation \eqref{dif non-d specific humidity}, we can write the water budget (column integrated moisture equation) as,
\begin{equation}
  \int^1_0 \Big(Ra^{-1/2}\lap q - \div (\bu q) \Big) dz = \int_{q > q_s}\frac{{q}-{q_s}}{{\tau}}dz \equiv P,
  \label{water budget} \\
\end{equation}
where z is the height, we have defined the precipitation term ($P$) to be the column integral of the moisture sink associated with condensation, with condensation defined as,
\begin{equation*}
    C \equiv \gamma \frac{q-q_s}{\tau} \mathcal{H}(q-q_s).
\end{equation*}
Similarly, the energy budget (column integrated buoyancy equation) can be written as,
\begin{equation}
  r - \int^1_0 \Big(Ra^{-1/2}\lap b - \div (\bu b) \Big) dz = \gamma \int_{q > q_s}\frac{{q}-{q_s}}{{\tau}}dz = \gamma P.
  \label{energy budget} \\
\end{equation}
It is common in climate change studies to separate the horizontal and vertical components of the budgets. We define the (diffusive) evaporation as,
\begin{equation*}
    E \equiv - Ra^{-1/2}\Bigg(\frac{\pa q}{\pa z} \bigg\rvert_{z = 0} - \frac{\pa q}{\pa z} \bigg\rvert_{z = 1}\Bigg),
\end{equation*}
i.e. the sum of the moisture source at the bottom boundary, and the moisture sink at the top boundary, due to diffusion. We also define the (diffusive) sensible heat flux as,
\begin{equation*}
    F_\textsf{sh} \equiv -Ra^{-1/2}\Bigg(\frac{\pa T}{\pa z} \bigg\rvert_{z = 0} - \frac{\pa T}{\pa z} \bigg\rvert_{z = 1}\Bigg) = -Ra^{-1/2}\Bigg(\frac{\pa b}{\pa z} \bigg\rvert_{z = 0} - \frac{\pa b}{\pa z} \bigg\rvert_{z = 1}\Bigg).
\end{equation*}
Equations \eqref{water budget} $\&$ \eqref{energy budget} can then be written as,
\begin{equation}
  P = E + \int^1_0 \Big(Ra^{-1/2}\lap_h q - \nabla_h \bcdot (\bu q)\Big) dz,
  \label{hor water budget} \\
\end{equation}
\begin{equation}
  \gamma P = r - F_{\textsf{sh}} - \int^1_0 \Big(Ra^{-1/2}\lap_h b - \nabla_h \bcdot (\bu b)\Big) dz. \label{hor energy budget} \\
  %\approx r + Ra^{-1/2} \big(1 + F_\textsf{ssh}\big).
\end{equation}
By integrating equations \eqref{hor water budget} $\&$ \eqref{hor energy budget} in the horizontal, and applying the divergence theorem and periodic horizontal boundary conditions, the domain averaged budgets simplify to 

\begin{equation}
  \overline{P} = \overline{E}, \label{water bs budget}
\end{equation}
\begin{equation}
  r = \gamma \overline{P} + \overline{F_{\textsf{sh}}}, \label{energy bs budget}
\end{equation}
where ${\overline{\:\cdot\:}}$ denotes the horizontal average of a quantity. The simplified water budget implies that the precipitation (moisture sink) is balanced by the evaporation (moisture source from diffusion), whilst the energy budget implies that the radiative cooling is balanced by the sum of heating associated with the precipitation ($\gamma P$) and the sensible heat flux (heat source from diffusion). Assuming the fluxes of buoyancy and moisture are small at the top of the domain (relative to the surface fluxes), the energy budget (approximately) relates the radiative cooling to the surface fluxes:
\begin{equation}
    r \approx \gamma \overline{E}_{\textsf{surf}} + \overline{F_{\textsf{ssh}}},
\end{equation}
Where $E_{\textsf{surf}} = -Ra^{-1/2}{\pa q}/{\pa z} (z = 0)$ is the surface evaporation, and $F_{\textsf{ssh}} = -Ra^{-1/2}{\pa T}/{\pa z} (z = 0)$ is the surface sensible heat flux.
\\
\newline
Note that, as $Ra \rightarrow \infty$, $E, F_{\textsf{sh}}\rightarrow 0$, so equation \eqref{water bs budget} $\Rightarrow P \rightarrow 0$, and the simplified energy budget equation \eqref{energy bs budget} becomes inconsistent unless $r \rightarrow 0$ as $Ra \rightarrow \infty$, indicating that (steady) equilibrium cannot exist in this limit.

\subsection{Moist Stability}
\begin{figure}
\begin{center}
  \includegraphics[width=0.6\textwidth]{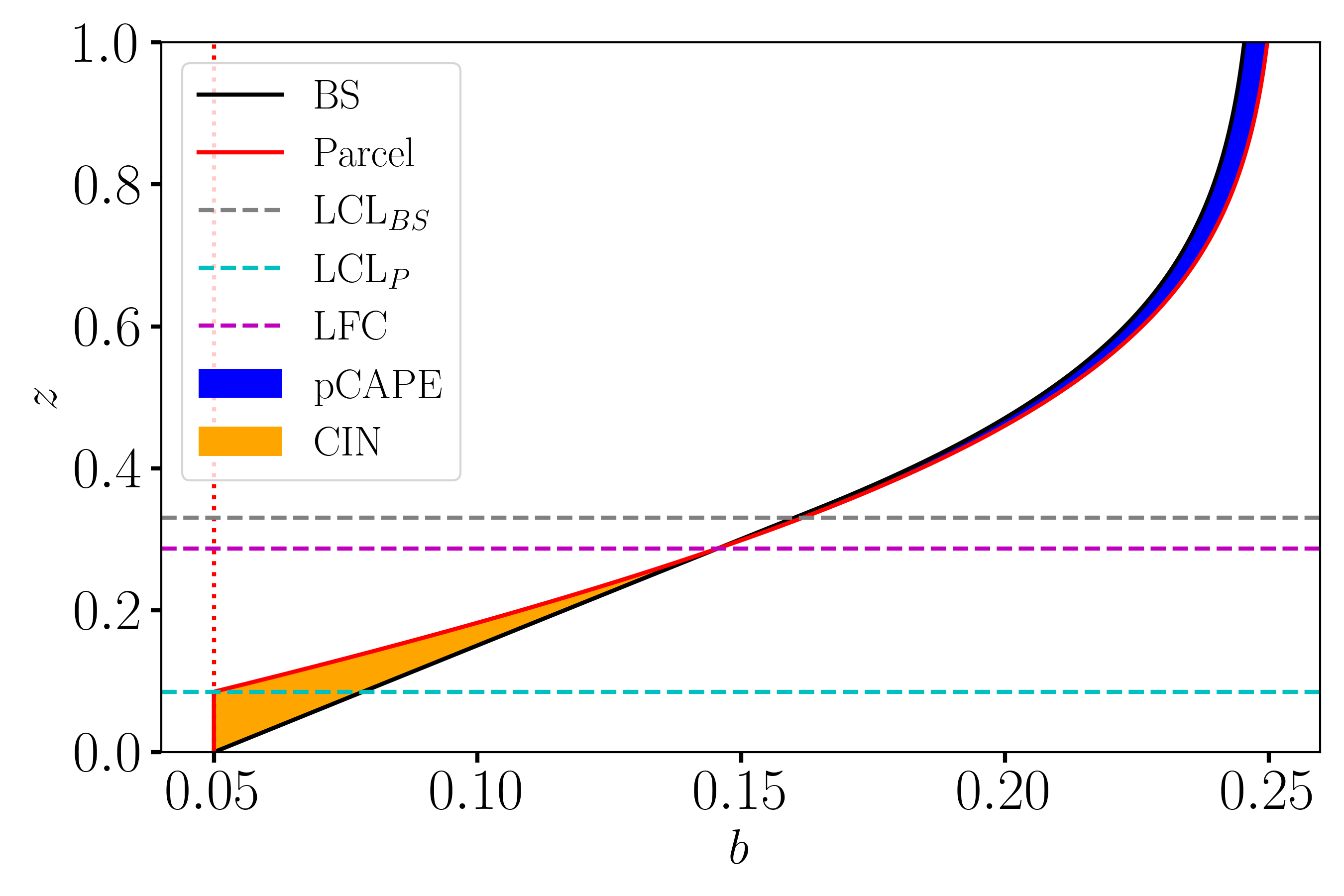}
  \caption{CAPE calculation schematic. The black line shows the basic state buoyancy curve, with the grey dashed line marking the basic state LCL. The red line marks the buoyancy profile of a parcel, the dashed cyan line marks the LCL of the parcel, and the dashed magenta line corresponds to the LFC, where the buoyancy of the parcel and the basic state are equal. The area of the blue shaded region is the positive CAPE, and the area of the orange shaded region gives the CIN.}
\label{fig:CAPE}
\end{center}
\end{figure}

To understand the moist stability of the atmosphere, it is typical in atmospheric science and meteorology to use a parcel argument \citep{Vallis_2017}. Consider a moist parcel of air. In the absence of diffusion and radiative cooling, the moist static energy of the parcel is conserved ($Dm/Dt = 0$). If the parcel of air is unsaturated ($q < q_s$), we rise the parcel along the dry adiabat ($b$ is conserved, so $dT/dz = -\Gamma_d = -g/c_p$) also conserving the humidity of the parcel (equation \eqref{dif non-d specific humidity} for $q<q_s$ and no diffusion), until the parcel becomes saturated at the (parcel) lifting condensation level (LCL), where $q = q_s$. We then rise the now saturated parcel along the moist pseudoadiabat, where $m$ is conserved and $q = q_s(T)$. The moist psuedoadiabatic buoyancy and temperature profiles of the atmospheric parcel (and the moist adiabatic lapse rate, $\Gamma_m = -dT_m/dz$) can be determined by solving the equation $m = b(T) + \gamma q(T) = m(0)$ as a function of height.
\\
\newline
We define the buoyancy and temperature of parcel using $b_p$ and $T_p$ respectively, and that of the environment with $b_E$ and $T_E$. An environmental profile can be defined as absolutely unstable, if the parcel satisfies $T_p \geq T_E \Rightarrow b_p \geq b_E$ at each height, $z$. In this case, the air parcel would be more buoyant (warmer) than its environment everywhere, and so would rise (or convect) freely. If instead the parcel is less buoyant (cooler) than its environment up to a height $z= \text{LFC} < H$, and more buoyant (warmer) than its environment for $\text{LFC} \leq z \leq \text{LNB}$, we can describe the atmosphere as conditionally unstable: for the parcel to convect freely (up to $z=\text{LNB}$, the level of neutral buoyancy), the parcel must be lifted (through turbulent or mechanical lifting) to the level of free convection (LFC). If the parcel is less buoyant (cooler) than its environment ($T_p \leq T_E \Rightarrow b_p \leq b_E$) at all heights $z$, the environment is absolutely stable. Note that conditional stability of a profile is no longer a local measure, because finite amplitude displacements may be needed to rise a parcel above it's LFC in order to release energy\footnotemark. An example of a conditionally unstable (basic state) environment is shown in Figure \ref{fig:CAPE}.
\footnotetext{Different definition of conditional instability in the American Meteorological Society Glossary: ``The state of a layer of unsaturated air when its lapse rate of temperature is less than the dry-adiabatic lapse rate but greater than the moist-adiabatic lapse rate''. For saturated upper profiles, the two definitions of conditional instability are equivalent, but not generally \citep{AMS_gloss}}
\vspace{6pt}
\\
\newline
To quantify the conditional instability of the environment, we use Convective Available Potential Energy (CAPE). Figure \ref{fig:CAPE} shows a schematic of how CAPE is calculated. The area of the orange region in Figure \ref{fig:CAPE} is the convective inhibition (CIN), where the parcel is less buoyant than its environment. Above the LFC, the area of the blue region is the positive CAPE (pCAPE), where the parcel is more buoyant than its environment. We additionally define (net) $\text{CAPE} = \int^{H}_0 (b_p-b_{BS})dz = \text{pCAPE}-\text{CIN}$. The positive CAPE can be thought of as the total amount of energy that can be released (by convection) in the system, and according to the parcel argument, positive CAPE is a necessary condition for the system to be able to release energy. However it does not tell us about convective onset in the system, i.e. when the system will overcome diffusion, which is determined by a linear stability analysis.

\subsection{Non-Dimensionalisations for Conditionally Unstable Atmospheres}
\label{CAPE non-dim}
The buoyancy non-dimensionalisation introduced in \cite{vallis2019simple} scales buoyancy with $[B] = g \Delta T / c_p$, where $\Delta T = T(0) - T(H)$. The boundary conditions on temperature (or buoyancy) and humidity were fixed at both the top and bottom boundaries in the \cite{vallis2019simple} study, so the temperature difference can be chosen independently of the humidity, as an external parameter. The temperature difference associated with our conditionally unstable (basic state) environments is always stable to dry air motion ($db/dz > 0 \text{ or } b(H) > b(0)$). Therefore this measure of buoyancy relates to the stabilising effects of warming of air as it descends with constant $b$. It has no dependence on the moisture in the system, and therefore it does not capture any changes in instability associated with variations in moist convective conditions. In the conditionally unstable system, instabilities are instead driven by the buoyancy of saturated air parcels rising above their LFC. We present two different scalings for $[B]$ (and $[L]$) which are more appropriate for (moist) conditionally unstable atmospheres. 
%Moist stability of the system is dependent on both $b$ (or $T$) and $q$, and so here we introduce a non-dimensionalisation based on the pCAPE of the environment, a measure of the moist instability of the environment.
 % Previous work on Bretherton study, and CAPE for the dry system.
\subsubsection{Bretherton Non-Dimensionalisation}
\cite{Bretherton_1988} introduced a moist Rayleigh number that captures some effects of conditional instability. Though his model differs from the Rainy-B\'enard model (see \cite{Bretherton_1987} for further details), the moist Rayleigh number can be simply calculated for our system. \cite{Bretherton_1988} defined the moist Rayleigh number as,
\begin{equation*}
  Ra_m = N_m^2 = \frac{H^4(\Gamma - N_d^2)}{\pi^4\nu^2}
\end{equation*}
Where $N_d^2 = (b_E(H)-b_E(0))/H$ is the Brunt-Vaisala frequency of the environment, and $\Gamma \approx (b_p(H)-b_p(0))/H$ is ``the buoyancy generation per unit rise due to latent heating'' in ``the (saturated) adiabatic ascent of a moist air parcel''. Note that the \cite{Bretherton_1988} study uses the conditions $N_d^2 > 0 \Rightarrow b_E(H) > b_E(0)$ and $\Gamma > N_d^2 \Rightarrow b_p(H) > b_E(H)$ to define a conditionally unstable atmosphere. For our model, we can define the moist `Bretherton' Rayleigh number ($Ra_m$) by setting $[L] = H, \: [B] = b_p(H)-b_e(H), \: [t] = \sqrt{[L]/[B]}$, which leads to the expression,
\begin{equation}
  Ra_m = \frac{H^3(b_p(H) - b_E(H))}{\kappa\nu}.
  \label{Breth Ra}
\end{equation}
The expression of $Ra_m$ given in equation \eqref{Breth Ra} captures the ratio of conditional instability (quantified by the buoyancy difference between a parcel and its environment at $z=H$) to diffusion. However, the simple buoyancy scale ($[B] = b_p(H)-b_E(H)$) does not account for the curvature of the environmental and buoyancy profiles as shown in Figure \ref{fig:CAPE}. For this reason, we here introduce a non-dimensionalisation based on the pCAPE of the environment, which provides an alternative (more accurate) measure of the conditional instability of the environment.
\begin{figure}[h]
\begin{center}
  \includegraphics[width=0.49\textwidth]{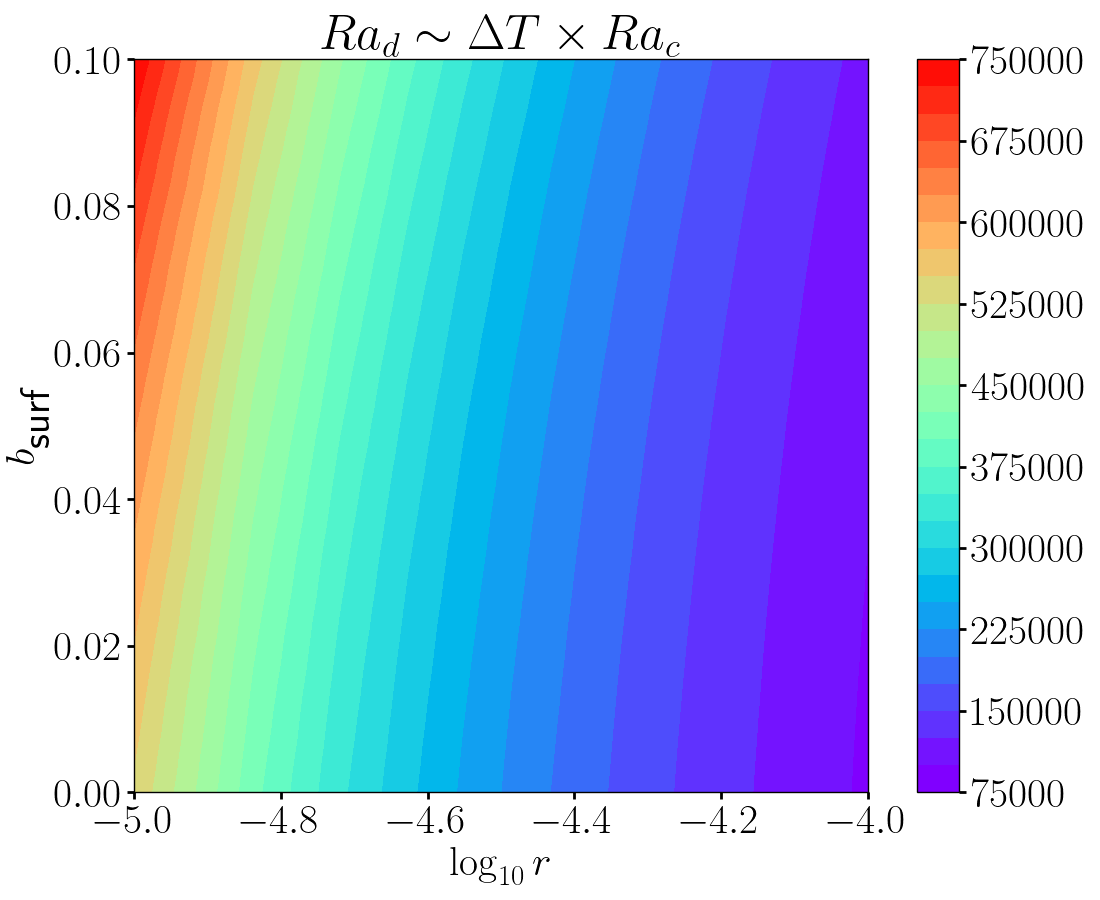}
  \includegraphics[width=0.49\textwidth]{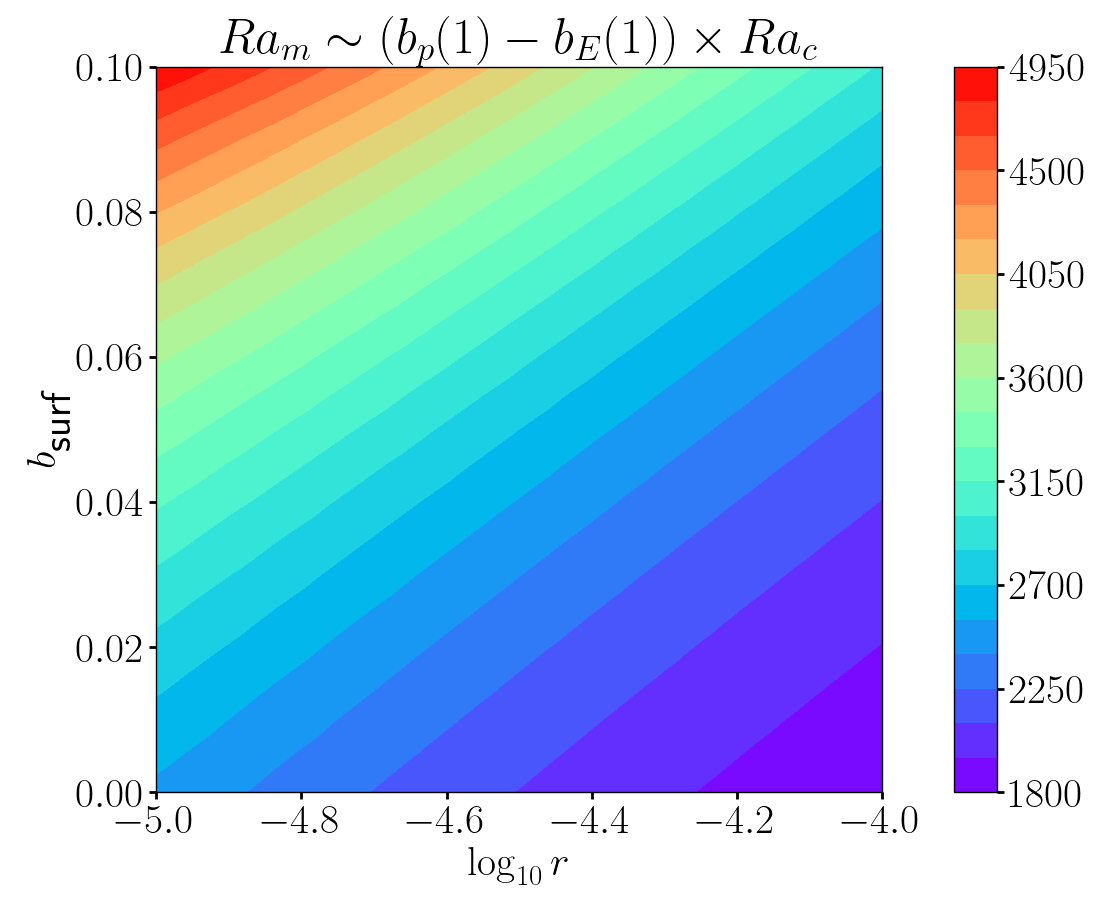}
  \\
  \includegraphics[width=0.49\textwidth]{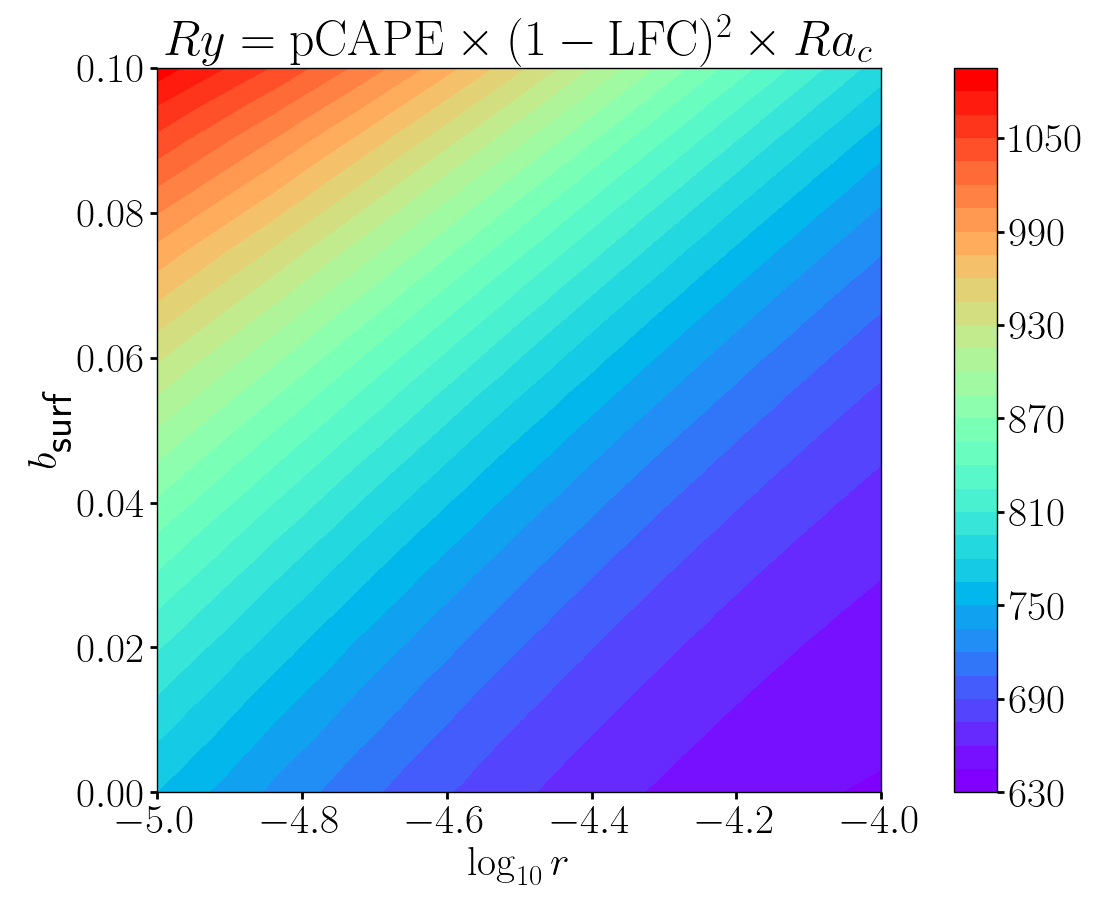}
  \includegraphics[width=0.49\textwidth]{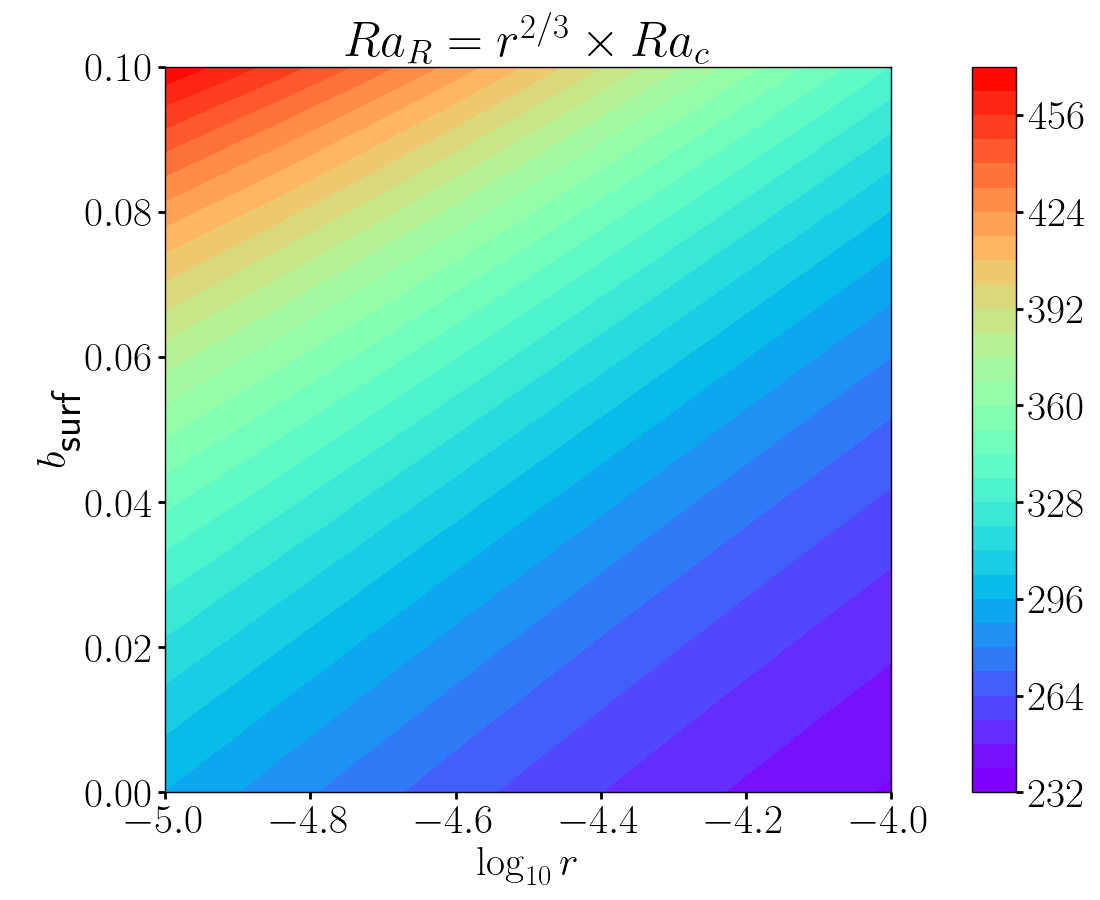}
  \caption{Critical dry Rayleigh number (top left), moist `Bretherton' Rayleigh number (top right), Rainy number (bottom left) and radiative Rainy Number (bottom right), as a function of the surface temperature increase and radiative cooling rate. The classical dry Rayleigh number shows variation of a factor of $\sim 10$ at criticality over the parameter space, while the moist `Bretherton' Rayleigh number varies by a factor of $\sim 2.75$, the radiative Rainy number varies by a factor of $\sim 2.03$ and the Rainy number varies by a factor of $\sim 1.75$.}
\label{fig:Rac_comp}
\end{center}
\end{figure}
\subsubsection{pCAPE Non-Dimensionalisation}
\label{CAPE nondim}
For a given set of boundary conditions, we first calculate the parcel buoyancy (or temperature) profile. Based on the environmental buoyancy (or temperature) profile, we can then calculate the pCAPE of the environment, which we use as a scale for the kinetic energy in the model, so that $[U] \sim \sqrt{\text{pCAPE}}$. Assuming that $b_p \geq b_e$ for $\text{LFC} \leq z \leq \text{LNB}$, we take the (CAPE) length scale to be $[L] = \text{LNB}-\text{LFC}$, which determines the buoyancy ($[B] \sim [U]^2/[L]$) and timescale ($[t] \sim [U]/[L]$) for the system. Note that for the environment shown in Figure \ref{fig:CAPE}, we set $\text{LNB}=H$, i.e. the domain height. The scales associated with this non-dimensionalisation are:
\begin{equation}
  [U] = \sqrt{\text{pCAPE}}, \qquad [B] = \frac{\text{pCAPE}}{\text{LNB}-\text{LFC}}, \qquad [L] = \text{LNB}-\text{LFC}, \qquad [t] = \frac{\text{LNB}-\text{LFC}}{\sqrt{\text{pCAPE}}}.
\end{equation}
Additionally, we set the pressure scale to be $[\Phi] = [U]^2 = \text{pCAPE}$. The non-dimensional momentum equation is then given by,
\begin{equation}
  \frac{D\bu}{Dt} = - \grad \phi + b \mathbf{k} + \bigg(\frac{Pr}{Ry}\bigg)^{\frac{1}{2}}\lap \bu,
\end{equation}
where,
\begin{equation}
    Ry \equiv \frac{\text{pCAPE}\times(\text{LNB}-\text{LFC})^2}{\nu\kappa}
    \label{Rainy number}
\end{equation}
is the ``Rainy'' number, which is a moist version of the Rayleigh number.
\\
\newline
The Rainy number represents the ratio of the destabilising effect of conditional instability (quantified by CAPE) to the stabilising effect of diffusion in the model, and therefore more faithfully corresponds to the role of Rayleigh number in dry classical Rayleigh-B\'enard convection. Figure \ref{fig:Rac_comp} shows the critical ``classical'' Rayleigh number, the critical moist Bretherton Rayleigh number, and the critical Rainy number across the climate change parameter space, as calculated by a linear stability analysis. Note that the linear stability analysis is discussed in detail in Section \ref{LSA Section}. The Rainy number varies much less than the ``classical'' (dry) critical Rayleigh number over the parameter space, as a result of including a measure of moist instability in the definition of the Rainy number. Additionally, the Rainy number varies less than the moist `Bretherton' Rayleigh number across the parameter space as a result of choosing CAPE (instead of $b_p(H)-b_E(H)$) to quantify conditional instability. The results in Figure \ref{fig:Rac_comp} therefore show that $Ry$ is more useful than $Ra_d$ and $Ra_m$ in categorising the behaviour of the moist conditionally unstable system. We need to know the basic state to be able to compute CAPE and hence $Ry$, however calculating CAPE is routine in meteorology and climate studies.
\\
\newline
We derive a radiative Rainy number in Appendix \ref{radiative Ry}, based on the surface flux of moist static energy, which is an alternative measure of conditional instability in the system. By quantifying the surface flux of $m$, the radiative Rainy number becomes relevant for climate studies focused on the response of moist convection to different surface fluxes. The CAPE based Rainy number given by equation \eqref{Rainy number} can be used to understand the water cycle intensification under climate change (observed previously in \cite{CP4}).

\section{Basic State Analysis}
\label{BS}
\begin{figure}[h]
    \centering
    \includegraphics[width=14.5cm]{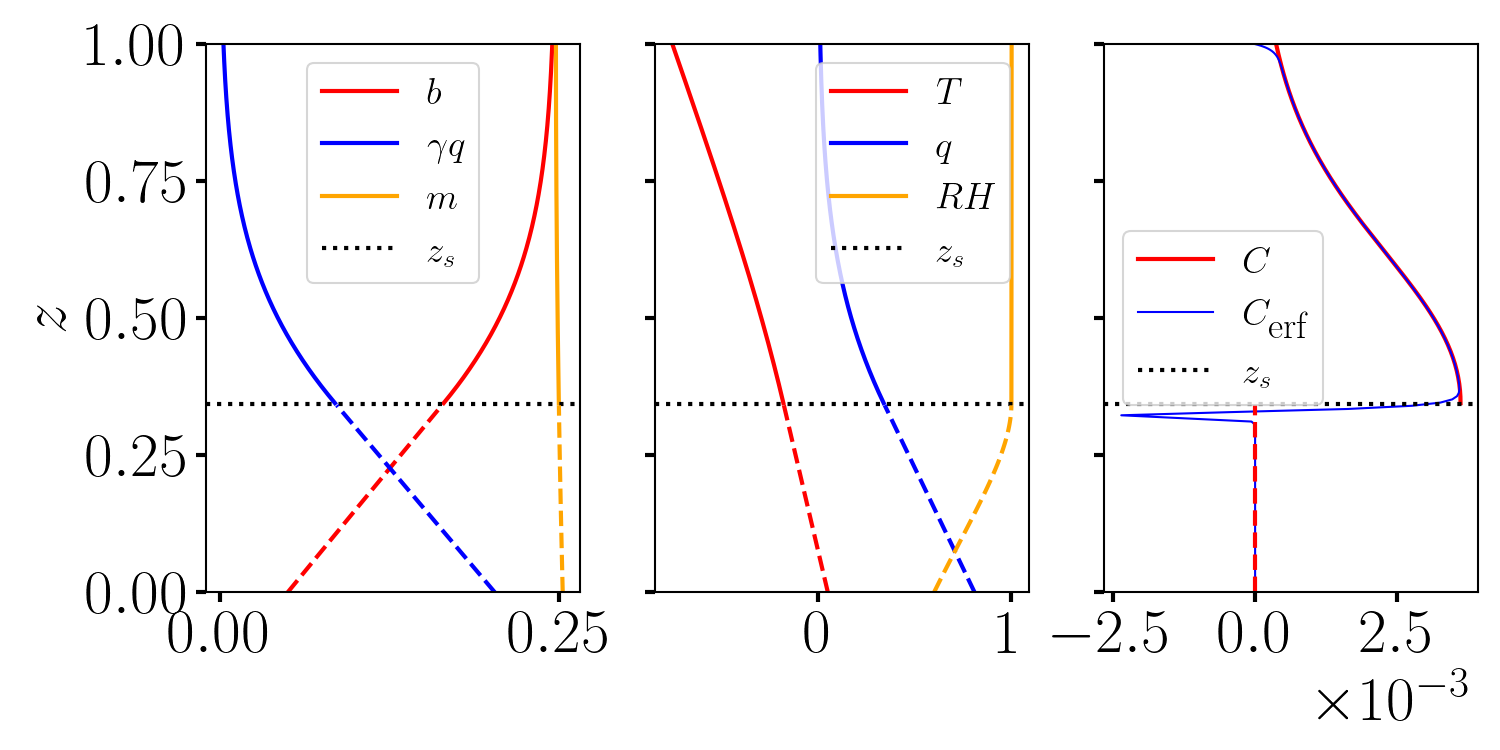}%
    \caption{Analytical basic state solution profiles for $RH_{\text{surf}} = 0.6$, $b_{\text{surf}} = 0.05$ and $r = 1\times10^{-5}$. The unsaturated part of the solution is given by the dashed lines, and the saturated part is given by the solid lines. The dotted red line marks the LCL. The numerical condensation profile ($C_{\textrm{num}}$) for a smooth approximation of $\mathcal{H}$ is also shown.}
\label{fig:bs_comp}
\end{figure}

To understand the behaviour of the model (under the dry adiabatic buoyancy non-dimensionalisation) with the boundary conditions specified in \ref{BC section}, a natural starting point is to calculate the basic state. \cite{Pauluis2010IdealizedState} examine steady state solutions to their simple model of moist convection, and in both \cite{vallis2019simple} and \cite{oishi2024_lsa} a ``drizzle solution'' of no motion is found for fixed temperature conditions, and saturated upper and lower boundary conditions. Here, the problem is similar to that of \cite{vallis2019simple} and \cite{oishi2024_lsa}, although the analysis is conducted for the idealised climate change set up of the Rainy-B\'enard model with radiation and moist psuedoadiabatic upper boundary conditions (outlined in \ref{BC section}), and moreover the solution is here found analytically (as in \cite{oishi2024_lsa}). Once the basic state solution is known, the stability of the system can be understood by calculating CAPE and the linear stability of the system (from the basic state). 
\\
\newline
The basic state we find is a $z$-dependent state of no motion ($\mathbf{u = 0}$), which is time-independent ($\partial/\partial t = 0$). For the boundary conditions outlined in \ref{BC section}, the basic state solution consists of a lower unsaturated region with an upper saturated region. Dropping the hats, the basic state equations (for $Pr=Sm=1$) are given by
\begin{align}
    \frac{d\phi}{dz} &= b,
    \label{basic momentum} \\
    \frac{d^2b}{dz^2} &= Ra^{1/2}\Bigg(-\gamma\frac{{q-q_s}}{\tau}\mathcal{H}({q}-{q_s})+r\Bigg),
    \label{basic buoyancy} \\
    \frac{d^2q}{dz^2} &= Ra^{1/2}\frac{{q-q_s}}{\tau}\mathcal{H}({q}-{q_s}),
    \label{basic specific humidity} \\
    {q_s} &= e^{\alpha({b} - {z})}. \label{basic sat specific humidity}
\end{align}
By examining the system of ODEs, one can see that,
\begin{equation*}
    \eqref{basic buoyancy}+\gamma\eqref{basic specific humidity} \implies \frac{d^2(b+\gamma q)}{dz^2} = \frac{d^2m}{dz^2} = Ra^{1/2}r.
\end{equation*}
Upon applying the boundary conditions in \ref{BC section}, the buoyancy can be related to specific humidity throughout the domain by
\begin{equation}
    m \equiv b+\gamma q = m_{\text{surf}} + Rz^2/2 - Rz,
    \label{MSE}
\end{equation}
where $R \equiv Ra^{1/2} r$. It follows that \eqref{basic sat specific humidity} can be written as
\begin{equation}
    q_s(q,z) = \exp\Big({\alpha(-\gamma q + m_{\text{surf}} + Rz^2/2 - (R+1)z)}\Big).
    \label{q_s(z,q)}
\end{equation}
The basic state solution can be determined by solving \eqref{basic specific humidity}, where $q_s(q,z)$ is given by \eqref{q_s(z,q)}. Due to the non-linear Heaviside function $\mathcal{H}$, \eqref{basic specific humidity} must be solved separately in the unsaturated and saturated regions. These solutions are matched at the lifting condensation level (LCL) where $z=z_s$ and $q^* = q(z_s) = q_s(z_s)$. The matching conditions used at the LCL are continuity of $b$, $q$, and their first derivatives. 
\\
\newline
Since $\epsilon \equiv \tau/Ra^{1/2}$ is small, we use an asymptotic approach for this analysis (expanding $q = q_0 + \epsilon q_1 + ...$). In the unsaturated region, we solve
\begin{equation}
    \frac{d^2q}{dz^2} = 0 \quad \Rightarrow \quad \frac{d^2q_0}{dz^2} + \epsilon \frac{d^2q_1}{dz^2} \approx 0.
\end{equation}
In the saturated region,
\begin{equation}
    \frac{d^2q}{dz^2} = \frac{q-q_s}{\epsilon} \quad \Rightarrow \quad q_0 = q_s(q_0) \quad \& \quad q_1 = \frac{d^2q_0/dz^2}{1+\alpha \gamma q_0},
\end{equation}
where we have used a first order Taylor expansion about $q_0$. The $\mathcal{O}(\epsilon)$ term is required to balance the diffusion of the $\mathcal{O}(1)$ solution. Note that the leading order solution has zero condensation and hence precipitation ($C(z) = 0 \Rightarrow P = 0$), since $q = q_s$ for $z \geq z_s$, and so we also need to consider the $\mathcal{O}(\epsilon)$ terms for the water budget equation \eqref{water bs budget} to be consistent. The analytical, asymptotic, basic state solution is given by $q = q_0 + \epsilon q_1 + ...$, where
\[
    q_0(z)= 
\begin{cases}
    q_{\text{surf}} + \Big(\frac{q^*-q_{\text{surf}}}{z_s}\Big)z,& \text{if } z < z_s\\
    \frac{1}{\alpha \gamma}W\Big(\alpha \gamma \exp \Big\{{\alpha(m_{\text{surf}} + \frac{R}{2}z^2 - (R+1)z)}\Big\}\Big),& \text{if } z \geq z_s
\end{cases}
\]
and,
\[
    q_1(z)= 
\begin{cases}
    0,& \text{if } z < z_s\\
    \frac{\alpha q_0(z)}{(1+\alpha \gamma q_0(z))^2}\Big(R+\alpha \left\{\frac{Rz-(R+1)}{1+\alpha \gamma q_0(z)}\right\}^2\Big),& \text{if } z \geq z_s.
\end{cases}
\]
Here $W$ is the Lambert-W function (the implicit solution of $W(x) \exp(W(x)) = x$ for any $x$). This solution was verified by comparison with a numerical solution to the non-linear boundary value problem (see Figure \ref{fig:bs_comp},\cite{oishi2024_lsa}, and supplementary material for further details of this comparison). Note that the moist pseudoadiabtic boundary conditions only hold to leading order: for the $q(1) = q_s(1)$ condition to hold in the $\mathcal{O}(\epsilon)$ solution, we require a boundary layer of $\mathcal{O}(\epsilon^{3/2})$, which is evident from the numerical profile of condensation in Figure \ref{fig:bs_comp}.
\\
\newline
Profiles of all other quantities can be deduced once $q(z)$ is known, by additionally using the solution $m(z)$ from equation \eqref{MSE}. The profiles are shown for a specific set of climate parameters in Figure \ref{fig:bs_comp}. Each of the profiles in Figure \ref{fig:bs_comp} show a lower unsaturated region (where $q<q_s$ or $RH < 1$), matched onto a saturated upper region (where $q \geq q_s$ or $RH \geq 1$), apart from the discontinuous condensation profile. Note that the basic state solution is saturated aloft, which differs from the real atmosphere, which is unsteady and nonlinear, but close to the moist pseudoadiabat with some conditional instability present. Figure 4 confirms how the upper profile is close to the moist pseudoadiabat, in contrast with \cite{vallis2019simple}, \cite{Agasthya_2024} and \cite{oishi2024_lsa} where the fixed upper temperature condition prevents this adjustment.

\subsection{Parameter Sensitivity of the Basic State Conditional Instability}
\begin{figure}[h]
\begin{center}
    \includegraphics[width=0.49\textwidth]{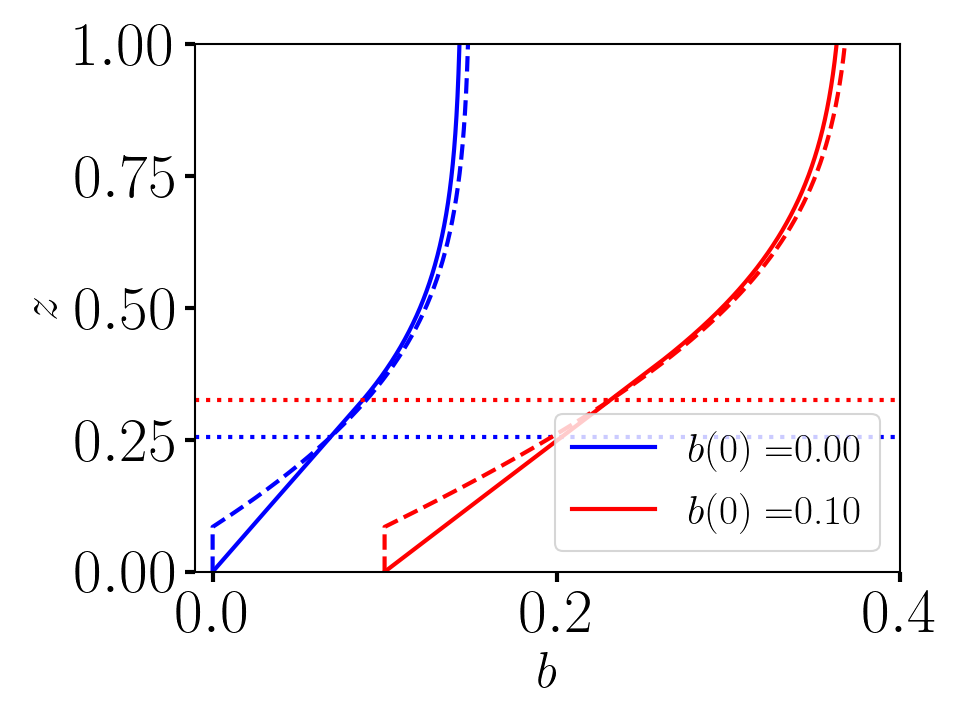}
    \includegraphics[width=0.49\textwidth]{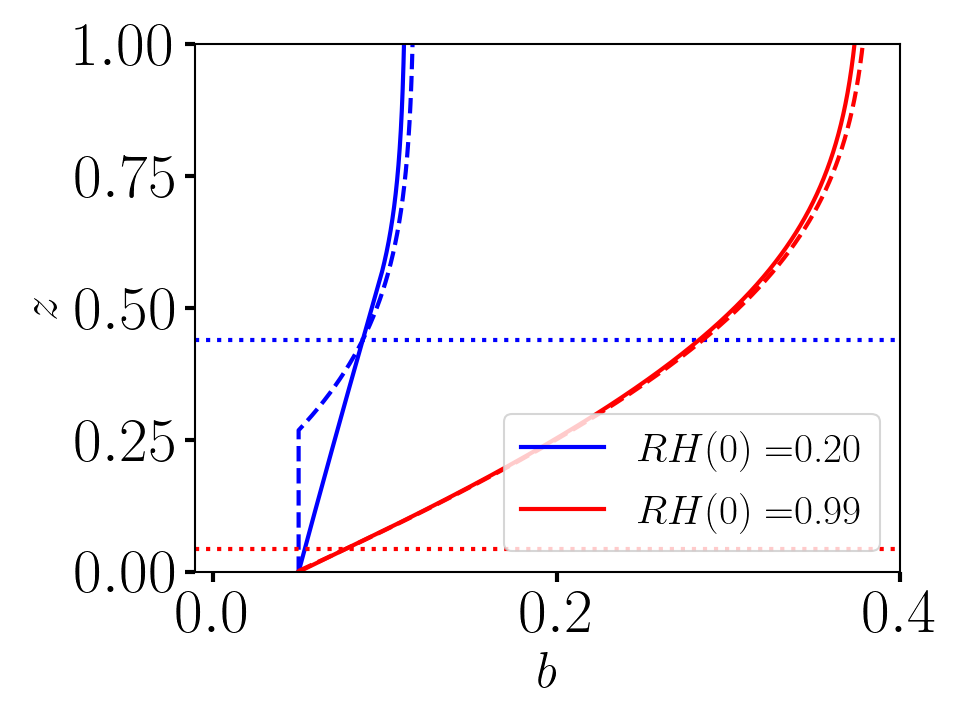}
    \\
    \includegraphics[width=0.49\textwidth]{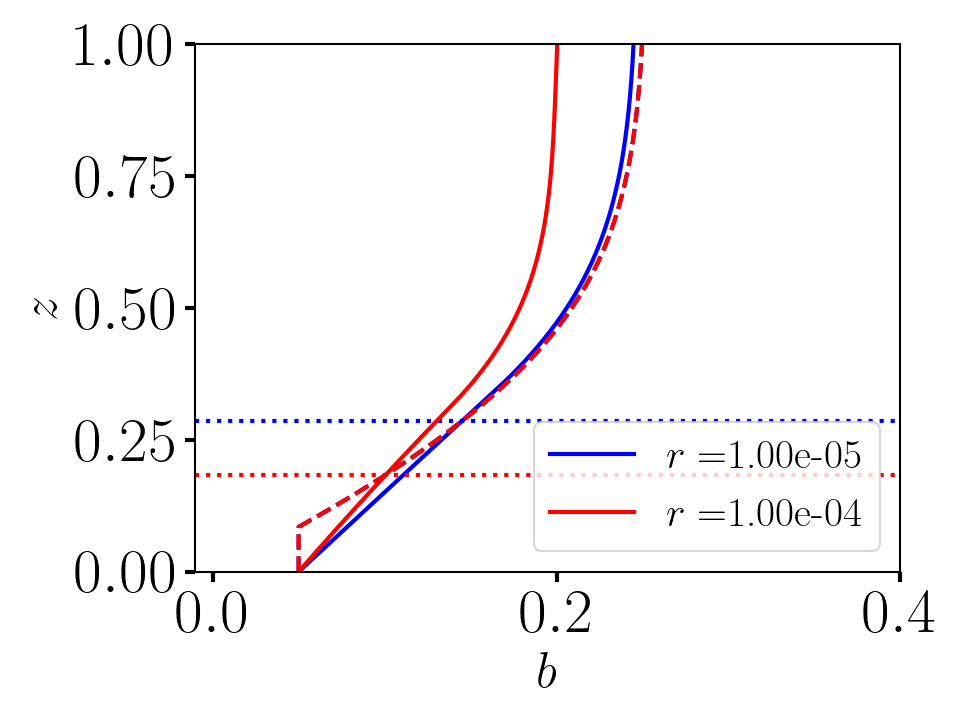}
    \includegraphics[width=0.49\textwidth]{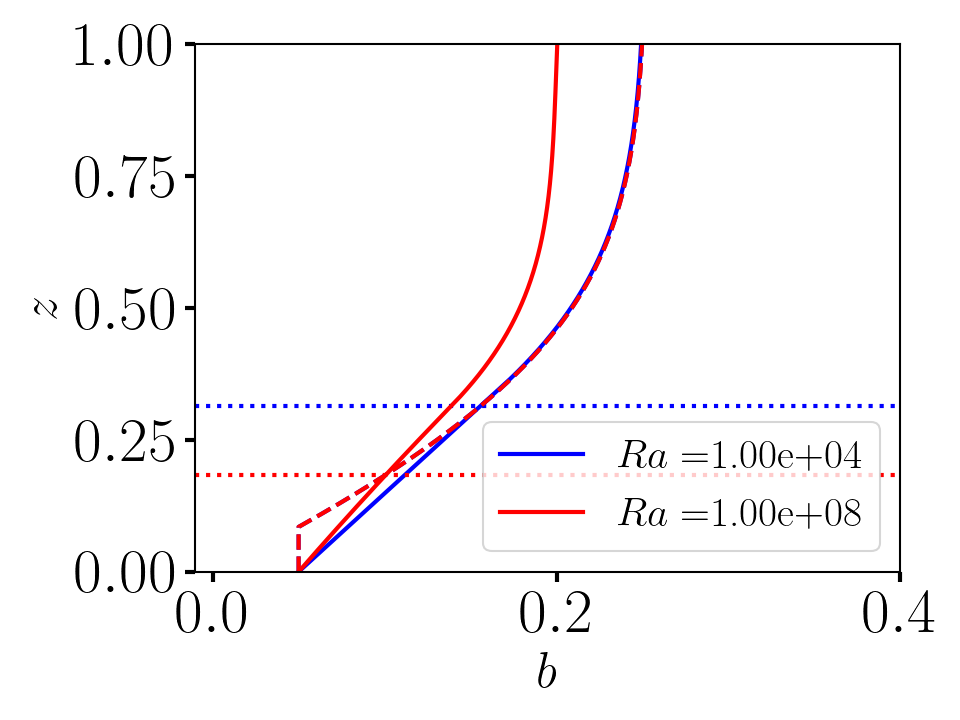}
    \caption{Parcel (dashed) and basic state (solid) profiles for varying: surface temperature (top left), surface relative humidity (top right), radiative cooling rate (bottom left) and Rayleigh number (bottom right). Unless otherwise stated (in the plot legend), the set of parameter values used are $b_{\textsf{surf}}=0.05,\: RH_{\textsf{surf}}=0.6,\: r = 10^{-5} \:\&\: Ra=10^6$.}
    \label{fig:bprofs}
\end{center}
\end{figure}

In this section, we analyse the sensitivity of the conditional instability of basic state environments to the climate parameters, by analysing buoyancy profiles and several diagnostics. We first illustrate the parameter dependence of the conditional instability by computing buoyancy profiles of both the basic state and a lifted surface parcel for different parameter values in Figure \ref{fig:bprofs}.
\\
\newline
The top left panel of Figure \ref{fig:bprofs} shows how an increase in surface temperature affects the conditional instability in the basic state environment. By comparing the (cooler) blue and (warmer) red profiles, we find that increasing the surface temperature (by $\sim 10K$, as in \cite{RCEMIP}) causes the LFC to increase, the CIN to increase and the pCAPE to decrease. We find that the changes in the conditional instability associated with increasing the surface temperature behave monotonically, at least for the range of parameters considered in our study. The reduction in conditional instability (characterised by a reduction in the net CAPE) with increasing surface temperature is due to the increased curvature of both the parcel and basic state buoyancy profiles, as a result of the presence of more moisture in the system associated with warmer surface temperatures. 
\\
\newline
The effect of surface relative humidity on the conditional instability is shown in the top right panel of Figure \ref{fig:bprofs}. We find that changing the surface relative humidity from $20\%$ to $99\%$ causes a significant decrease in the LFC and CIN, relative to the small increase in pCAPE. Note that in the limit $RH_{\textsf{surf}} \rightarrow 0$, the LFC $\rightarrow 0$ if the environment is dry unstable (as it would be for the basic state with dry adiabatic BCs at the top boundary), indicating that the dependence of the conditional instability (LFC, CIN and pCAPE) on the surface relative humidity does not behave monotonically.  
\\
\newline
The effect of increasing the radiative cooling and Rayleigh number on the conditional instability is illustrated by the bottom panels of Figure \ref{fig:bprofs}. Since the (dashed) parcel profile is independent of both $r \: \& \: Ra$, changes in conditional stability are due to changes in the basic state solution. Increases in $r$ and $Ra$ have the same (monotonic) effect on conditional instability: they both cause a decrease in the LFC, increase in the CIN and a significant increase in the pCAPE, relative to the changes in pCAPE caused by changing $b_{\textsf{surf}}$ or $RH_{\textsf{surf}}$.
\\
\newline
\subsubsection{Conditional Instability and Moisture Diagnostics}
To examine how the basic state changes under climate change (increasing both $r \:\&\: b_{\textsf{surf}}$), we can compute both conditional instability diagnostics (pCAPE, CIN, LFC $\& \: Ry$), and moist diagnostics ($P$,$z_s$). We define a `typical' climate change scenario, consistent with Figure 4 of \cite{SimpleSpectralModelsforAtmosphericRadiativeCooling}, by assuming that the surface relative humidity remains fixed (e.g. over an ocean surface), with the radiative cooling doubling from $\sim1 K/\textrm{day}$ to $\sim2 K/\textrm{day}$ (increasing $r$ from $1\times 10^{-5}$ to $2\times 10^{-5}$) when we increase the surface temperature by $\sim 10K$ (increasing $b_{\textsf{surf}}$ from $0.0$ to $0.1$). We consider the effects of the typical climate change scenario, and alternative climate change scenarios, on the system by computing conditional instability and moisture diagnostics with the surface relative humidity, for a fixed set of surface temperatures and radiative cooling values, and additionally fixed $Ra=10^6$. We plot the diagnostics against surface relative humidity since the conditional instability diagnostics have a non-monotonic relationship with $RH_{\textsf{surf}}$, but vary monotonically with $b_{\textsf{surf}},\:r\:\&\:Ra$.
\begin{figure}[h]
    \centering
    \includegraphics[width=0.95\textwidth]{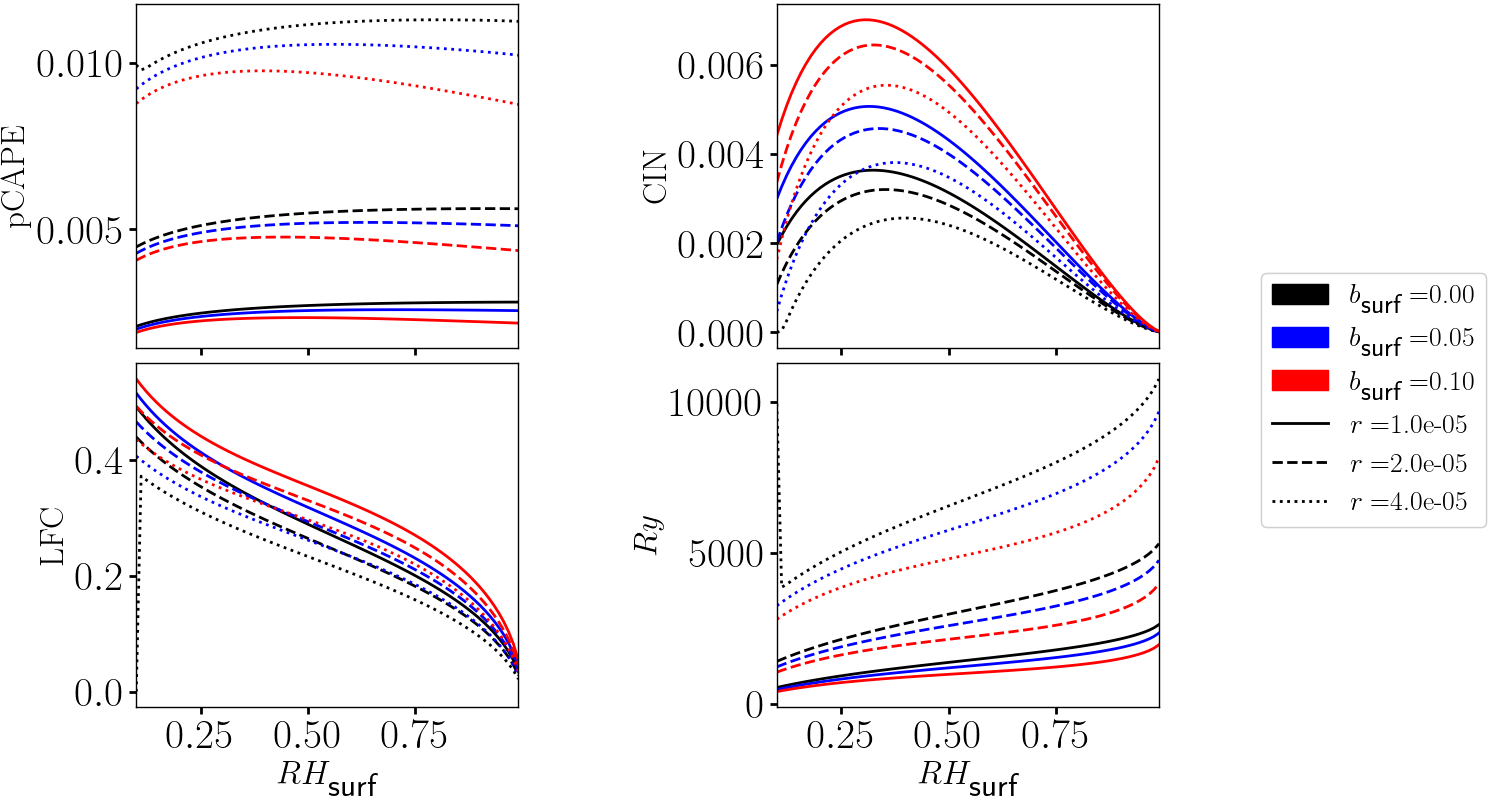}%
    \caption{Conditional instability diagnostics (for the basic state) as a function of the surface relative humidity. The different line colours represent different values of surface temperature increase, and the different line styles represent different radiative cooling rates. The Rayleigh number is held fixed at $Ra=10^6$.}
\label{fig:diags_CAPE}
\end{figure}
\begin{figure}[h]
    \centering
    \includegraphics[width=0.90\textwidth]{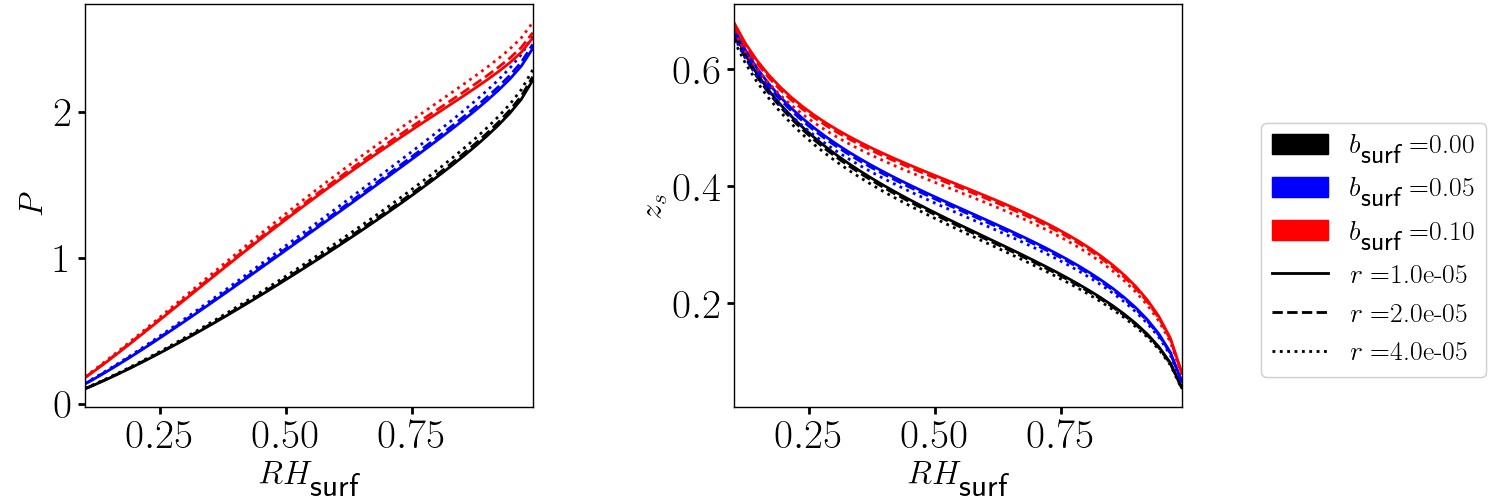}%
    \caption{Basic state moisture diagnostics (precipitation, $P$, and the LCL, $z_s$) as a function of the surface relative humidity. The different line colours represent different values of surface temperature increase, and the different line styles represent different radiative cooling rates. The Rayleigh number is held fixed at $Ra=10^6$.}
\label{fig:diags_moist}
\end{figure}
\\
\newline
The analysis in Section \ref{CAPE non-dim} showed that the Rainy number is a useful parameter to understand the behaviour of the basic state environment: if $Ry \sim \text{pCAPE} \times (1-\text{LFC})^2\times Ra$ increases, we expect increased levels of (moist conditional) instability in our system. The Rainy number is shown in the bottom right panel of Figure \ref{fig:diags_CAPE}. In general we see that $Ry$ increases with $RH_{\textsf{surf}}$ (for fixed $r$ and $b_{\textsf{surf}}$). For varying $RH_{\textsf{surf}}$, changes in $Ry$ are dominated by changes in the LFC rather than changes in pCAPE. The kink in $Ry$ along the black dotted line (as $RH_{\textsf{surf}} \rightarrow 0$) indicates the transition from a moist conditionally unstable atmosphere (where $\text{LFC, CIN} > 0$), to a (dry) absolutely unstable atmosphere ($b_p \geq b_E \:\forall z\in [0,1] \:\Rightarrow \text{LFC, CIN}=0$). We find that increasing $b_{\textsf{surf}}$ (with $RH_{\textsf{surf}}$ and $r$ fixed) causes $Ry$ to decrease, as a result of both the LFC increasing and the pCAPE decreasing. Similarly, $Ry$ increases with $r$ (assuming $b_{\textsf{surf}}$ and $RH_{\textsf{surf}}$ remain fixed), as a result of both the LFC decreasing and the pCAPE increasing with increasing radiative cooling. Figure \ref{fig:diags_moist} shows how the moisture in the basic state changes under climate change, characterised by precipitation term ($P$, defined in Section \ref{moiststab}) and the LCL ($z_s$) of the basic state: $P$ increases in response to increases in all of the climate parameters ($r$, $b_{\textsf{surf}} \: \& \: RH_{\textsf{surf}}$), however $z_s$ decreases with increasing $r$ and $RH_{\textsf{surf}}$, but increases with increasing $b_{\textsf{surf}}$. Considering both Figures \ref{fig:diags_CAPE} $\&$ \ref{fig:diags_moist} together, it follows that the $r$ is the most important parameter (for fixed $Ra$) in affecting the degree of stability in the model (characterised by $Ry$), however $b_{\textsf{surf}}$ and $RH_{\textsf{surf}}$ are key parameters for affecting the amount of moisture is in the system, since changes in $r$ does not affect the value of $P$ or $z_s$ nearly as much as changes in $RH_{\textsf{surf}}$ or $b_{\textsf{surf}}$.
\\
\newline
The response of the moist instability in the system to different climate change scenarios can be considered by examining the response of $Ry$ to an increase in both the radiative cooling rate and the surface temperature, together. Recall that the typical climate change scenario involves increasing both $r$ from $1\times 10^{-5}$ to $2\times 10^{-5}$ and $b_{\textsf{surf}}$ from $0.0$ to $0.1$, assuming that $RH_{\textsf{surf}}$ remains constant. In Figures \ref{fig:diags_CAPE} and \ref{fig:diags_moist}, the typical climate change scenario therefore involves going from the black solid line to the red dashed line, keeping the surface relative humidity fixed. Under the typical climate change scenario, Figure \ref{fig:diags_CAPE} shows that $Ry$ increases, and Figure \ref{fig:diags_moist} shows $P$ and $z_s$ also increase (for the basic state). both $P$ and $z_s$ increase indicating that climate change causes increased levels of moist instability (in the basic state). Under the typical climate change scenario, the increased moist instability is dominated by effect of $r$ on $Ry$, whereas the increased levels of moisture are primarily associated with the effect of $b_{\textsf{surf}}$ on $P$. The basic state analysis therefore indicates that there will be more intense moist convection under climate change, due to increased moist instability and precipitation (over a smaller vertical region) in the basic state environment.

\section{Linear Stability Analysis}
\label{LSA Section}
\begin{figure}
\begin{center}
  \includegraphics[width=0.59\textwidth]{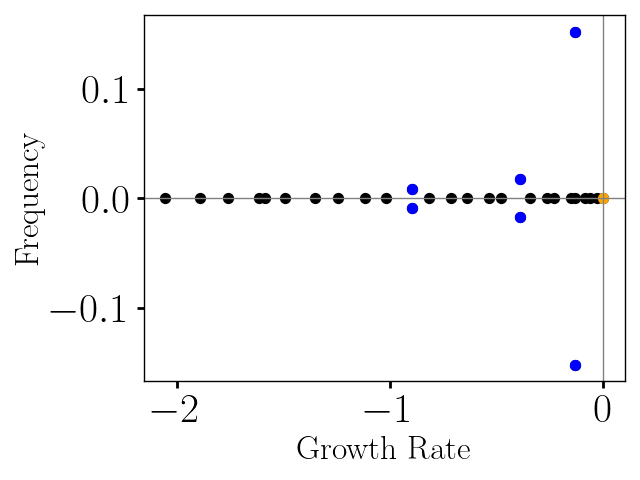}
  \caption{Eigenvalue spectra at criticality $Ra= Ra_c, \: k = k_c$. The blue dots show damped oscillatory modes, and the black dots show purely decaying modes. The results are shown for the parameter values: $RH_{\textsf{surf}} = 0.6$, $b_{\textsf{surf}} = 0.05$ and $r = 1\times10^{-5}$ and $Ra = Ra_c = 7.80\times10^5$.}
\label{fig:base_spectra}
\end{center}
\end{figure}

Analysis of linear stability can been used to provide initial insight into numerical (non-linear) simulations. In the textbook \cite{1961Chandrasekhar}, linear stability is analysed for dry classical Rayleigh-Bénard convection, to find the critical Rayleigh number and the corresponding wavenumber. For Rayleigh numbers greater than the critical Rayleigh number, convection sets in. \cite{vallis2019simple} found that for Rayleigh numbers that are just slightly greater than critical Rayleigh number, a series of steady convective updraughts are produced in the non-linear simulations. For dry Rayleigh-B\'enard convection, as the Rayleigh number is increased the regime of convection shifts from conductive ($Ra < Ra_c$) to steady (for $Ra > Ra_c$), to periodic, before becoming turbulent (see \cite{Waleffe_2015}). Finding the critical Rayleigh number (which varies with the parameters) allows one to compare non-linear simulations for different parameter values by running these simulations at the same supercriticality (e.g. $5\times Ra_c$). The critical wavenumber also provides an initial indication of how the width of the plumes may respond to change in the parameters. Therefore, we perform a linear stability analysis for the Rainy-B\'enard model here.
\\
\newline
We set up the eigenvalue problem by perturbing the basic state, $b = \Bar{b} + b'$, and assuming the linear wave ansatz, $b' = \hat{b}(z)e^{i(k_x x + k_y y - \sigma t)}$. The model equations can be reduced to the eigenvalue problem,
\begin{align}
    -i\sigma \hat{u} &= -ik_x \hat{\phi} + \frac{1}{Ra^{1/2}}\Big(\frac{d^2}{dz^2}-k^2\Big)\hat{u}, 
    \label{wave u} \\
    -i\sigma \hat{v} &= -ik_y \hat{\phi} + \frac{1}{Ra^{1/2}}\Big(\frac{d^2}{dz^2}-k^2\Big)\hat{v}, 
    \label{wave v} \\
    -i\sigma \hat{w} &= -\frac{d\hat{\phi}}{dz} + \:\hat{m} -\gamma \hat{q} + \frac{1}{Ra^{1/2}}\Big(\frac{d^2}{dz^2}-k^2\Big)\hat{w}, 
    \label{wave w} \\
    -i\sigma \hat{m} &= -\hat{w}\frac{d\Bar{m}}{dz} +  \frac{1}{Ra^{1/2}}\Big(\frac{d^2}{dz^2}-k^2\Big)\hat{m}, 
    \label{wave b} \\
    -i\sigma \hat{q} &= -\hat{w}\frac{d\Bar{q}}{dz} - \frac{\hat{q}-\alpha\Bar{q_s}\hat{b}}{\tau}\Big\{\mathcal{H}(\Bar{q}-\Bar{q_s}) + (\bar{q}-\bar{q_s})d\mathcal{H}(\bar{q}-\bar{q_s}) \Big\} + \frac{1}{Ra^{1/2}}\Big(\frac{d^2}{dz^2}-k^2\Big)\hat{q}, 
    \label{wave q} \\
    -\frac{d\hat{w}}{dz} &= ik_x \hat{u} + ik_y \hat{v},
    \label{wave incomp}
\end{align}
with boundary conditions
\begin{align}
    &\hat{m}(0) = 0, \quad \hat{q}(0) = 0, \quad \mathbf{\hat{u}}(0) = \mathbf{0}, \quad 
    \\
    &\frac{\hat{dm}}{dz}(1) = 0, \quad \Big\{1+\alpha \gamma \overline{q_s}(1)\Big\}\hat{q}(1) - \alpha \overline{q_s}(1) \hat{m}(1) = 0, \quad \mathbf{\hat{u}}(1) = \mathbf{0}.
    \label{wave BCs}
\end{align}
Note that $k^2 \equiv k_x^2 + k_y^2$, and, a first order Taylor expansion is used to express $q_s' \approx \alpha \bar{q}_s b'$. In this formulation, we have replaced $\hat{b}$ with $\hat{m}$ to implement the upper moist pseudoadiabatic boundary conditions. The second boundary condition in equation \eqref{wave BCs} is equivalent to $q = q_s$. We solve the eigenvalue problem numerically using the Dedalus framework \citep{Dedalus}. The eigenvalue problem given by equations \eqref{wave u} - \eqref{wave BCs} is for a smooth approximation of the Heaviside function, which allows $\mathcal{H}$ to be Taylor-expanded (about $\bar{q}-\bar{q_s}$). We approximate $\mathcal{H}(x) = \frac{1}{2}(1+\text{erf}(kx))$ with $k=10^3$ so that, $d \mathcal{H}(x) = \frac{k}{\sqrt{\pi}}e^{-(kx)^2}$. Including this $d\mathcal{H}$ term is essential for accurately predicting the onset of convection for non-linear simulations which use a smooth approximation of the Heaviside function.
\\
\newline
The first step in the solving the EVP is to calculate the numerical basic state solution (for smooth $\mathcal{H}$). The numerical basic state solution is obtained by using Dedalus to solve the non-linear boundary value problem (given by equations \eqref{basic momentum} - \eqref{basic sat specific humidity}), with the analytical solution as an initial guess (to speed up convergence).
\\
\newline
To compute the critical Rayleigh number ($Ra_c$), we first calculate the basic state for smooth $\mathcal{H}$ (numerically) for a given Rayleigh number ($Ra$). The numerical basic state solution is obtained by using Dedalus to solve the non-linear boundary value problem (given by equations \eqref{basic momentum} - \eqref{basic sat specific humidity}), with the analytical solution as an initial guess (to speed up convergence). We then solve the eigenvalue problem for a range of wavenumbers, $k = \sqrt{k_x^2+k_y^2}$, and determine $k_{\max}$ which maximises the growth rate ($\Im(\sigma)$). We adjust $Ra$ iteratively, and repeat the above method, until the absolute value of the maximum growth rate is less than a tolerance, which we take to be $10^{-8}$. The pair $(Ra,k_{\max})$ which satisfies these conditions determines the critical Rayleigh number, $Ra_c$, and the critical wavenumber, $k_c$. 
\\
\newline
An example of the eigenvalue spectra for an unsaturated atmosphere ($RH_{\textsf{surf}} < 1$) at criticality is shown in Figure \ref{fig:base_spectra}. The maximum growing mode (yellow dot) has zero growth rate and frequency. Note that the damped oscillatory modes are only present for unsaturated cases: in the saturated limit ($RH_{\textsf{surf}} = 1$), there are no decaying oscillatory modes, i.e. all eigenvalues have zero frequency, indicating that the damped waves associated with the decaying oscillatory mode can only exist in the lower unsaturated region of the domain. The waves associated with the damped oscillatory mode are associated with dry gravity waves, and we derive an approximate form for their dispersion relationship in Section \ref{approx disp section}.

\subsection{Action of the Linear Perturbation}
In this section we examine the structure of the eigenvectors, the perturbations and their impact on the basic state, in terms of their effect on moisture and conditional instability. We compute the eigenvectors and perturbations corresponding to the fastest growing mode, at criticality ($Ra = Ra_c \;\&\; k=k_c$). Note that we define a (real) perturbation quantity, associated with an eigenvector, as for example $w' = \Re(\hat{w}(z) e^{ikx})$.
\begin{figure}[h]
\begin{center}
  \includegraphics[width=0.495\textwidth]{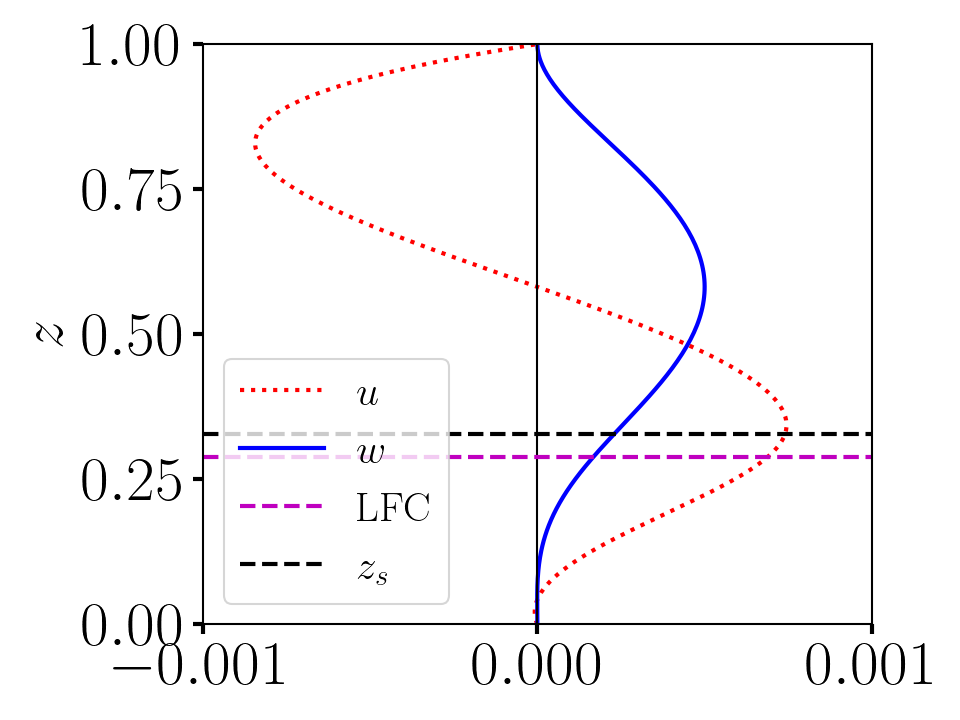}
  \includegraphics[width=0.495\textwidth]{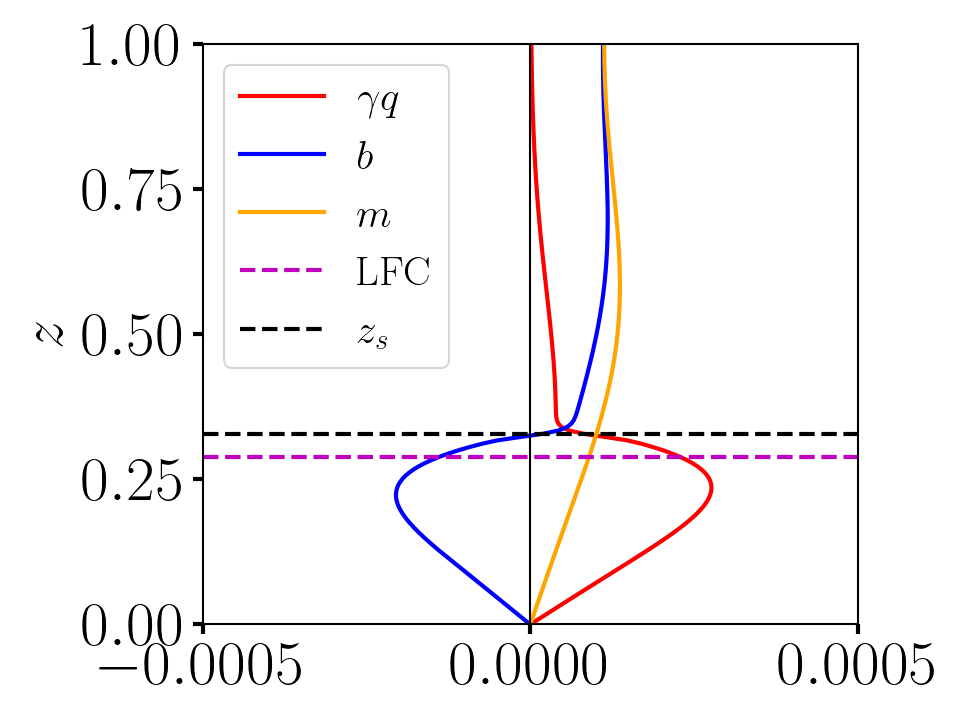}
  
  \caption{An example of normalised eigenvectors at criticality, $Ra=Ra_c \;\&\; k=k_c$, associated with the fastest growing mode. Note that the solid and dotted lines represent the real and imaginary parts of the eigenvectors, respectively. The left panel shows the imaginary part of $u$, and the real part of $w$. The right panel shows the (real) $b$, $\gamma q$ and $m$ eigenvectors. The dashed lines mark the basic state LCL ($z_s$, red) and the LFC (blue). Note that the normalisation is such that $w_i = b_i = q_i = u_r = 0$, so $u$ is $\pi/(2k_c)$ out of phase with $w,\: b,\: q\:\: \& \:\:m$, where $k_c$ is the critical wavenumber. The results are shown for the parameter values: $RH_{\textsf{surf}} = 0.6$, $b_{\textsf{surf}} = 0.05$ and $r = 1\times10^{-5}$ and $Ra = Ra_c = 7.80\times10^5$.}
\label{fig:base_evs}
\end{center}
\end{figure}
\\
\newline
Figure \ref{fig:base_evs} shows the vertical structure of the normalised eigenvectors. The eigenvectors are normalised such that $\max(w_r) = 1$ with $w_i = b_i = q_i = u_r = 0$. The eigenvector represents the perturbation in an updraft region (since $\max(w') > 0$). We can consider the perturbation in a subsiding region (where $\max(w') < 0$), by multiplying the eigenvectors in Figure \ref{fig:base_evs} by a factor of $-1$.
\\
\newline
The structure of the $\hat{b}$ and $\hat{q}$ eigenvectors in Figure \ref{fig:base_evs} reveals that the linear perturbation causes a moistening of updraft regions, and makes the updrafts more buoyant ($b' > 0$ for $z > z_s$). Assuming $x=0 \Rightarrow w' = \hat{w}, \; b' = \hat{b}, \; q' = \hat{q}$, etc., the relative humidity is given by $RH = q/q_s \approx (\bar{q} + q')/\big(\bar{q}_s(1 + \alpha b')\big) \Rightarrow RH' \approx (q'-\alpha \bar{q} b')/\bar{q}_s$. Below $z_s$, $b' < 0$ and $q' > 0 \Rightarrow RH' > 0$, so the linear perturbation moistens the updrafts lower (unsaturated) region. Computation of the perturbation relative humidity reveals that $RH' \geq 0$ for $z \geq z_s$ too, i.e. $q' \geq \alpha \bar{q}b'$ despite increase in $q_s$ due to warming. However the increase in $RH$ is much larger in the lower unsaturated region than it is in the upper saturated region. The moistening effect of the linear perturbation in updraft regions results in increased levels of condensation, and a decrease in $z_s$ in updraft regions. The linear perturbation has the opposite effect in subsiding regions, causing a drying effect which reduces the levels of condensation and increases $z_s$.
\\
\newline
To assess the effect of the linear perturbation on the conditional stability, we examine the structure of the perturbation $b'$. The parcel buoyancy profile remains the same, however the buoyancy profile of the environment changes with the linear perturbation. The $\hat{b}$ eigenvector in Figure \ref{fig:base_evs} shows the effect of the linear perturbation in an updraft region. For $z < z_s$, $b' = \hat{b} < 0$, so that the linear perturbation is making the environment less buoyant. Recalling Figure \ref{fig:CAPE}, this reduces the CIN in the lower region, and causes the LFC (where $b_p = b_E$) to decrease. There is a small vertical region above the LFC for which $b' < 0$, and so the pCAPE in the region is increasing. However, above $z_s$, the linear perturbation makes the environment more buoyant since $b' > 0$ here, which reduces the pCAPE. Therefore, the linear perturbation acts in updraft regions by making the lower levels more unstable by reducing the CIN, LFC and increasing the pCAPE in a small vertical region just above the LFC, however, it has a stabilising influence on the upper levels, causing a reduction in the overall pCAPE. The effect of the linear perturbation on the subsiding regions is opposite to that in the updraft regions: the linear perturbation acts by stabilising the lower levels up to just above the LFC, and destabilises the upper levels.
\\
\newline
The modal structure of the perturbation quantities over one wavelength is exhibited in Figure \ref{fig:base_modal}. Note that the eigenvectors are normalised with $\max(w_r) = 2 \times 10^{-3}$ to keep the perturbations small, relative to the basic state solution (to be consistent with Figure \ref{fig:base_cond}). The arrows show the circulation from the moister updraft regions to the drier subsiding regions. The perturbations $b' \; \& \; q'$ show sharp changes in gradient around the saturation point, $z_s$ (also shown in Figure \ref{fig:base_evs}). Note for this set of parameter values, in the updraft regions, there is a small region where $w' < 0$ close to the surface, although the magnitude of $w'$ is small compared to $\max (w')$ in the column. Figure \ref{fig:base_modal} shows that $w', \; b' \; \& \; q'$ are all in phase, with the circulation from the moister updraft regions to the drier subsiding regions associated with $u'$ being $\pi/(2k_c)$ out of phase with $w'$ as expected. 
\\
\newline
Figure \ref{fig:base_cond} shows the condensation term over one wavelength, produced by adding the enlarged linear perturbation to the basic state. We take $b = \bar{b} + b',\; q = \bar{q} + q'$ and compute $C = \gamma (q-q_s) \mathcal{H}(q-q_s),$ where $q_s = e^{\alpha b}$. Note that the fields of $b$ and $q$ look relatively unaffected by the addition of the linear perturbation, however the linear perturbation causes a significant change in the condensation term. The plots show two pockets around the LCL, where the condensation has been enhanced (the most) by the moistening effect of the linear perturbation in updraft regions. We can also see that the LCL has been perturbed further downwards at $x = 0$ than upwards at $x = \pi/k_c$. The enhanced levels of condensation around the LCL show that the linear perturbation causes the instability to stem from the LCL. The action can be clearly seen seen by running super critical simulations from the basic state plus the linear perturbation at a small amplitude (see supplementary material).

\begin{figure}[H]
\begin{center}
  %\includegraphics[width=0.69\textwidth]{figs/psipert.png}
  %\\
  %\includegraphics[width=0.69\textwidth]{figs/wpert.png}
  %\\
  \includegraphics[width=0.69\textwidth]{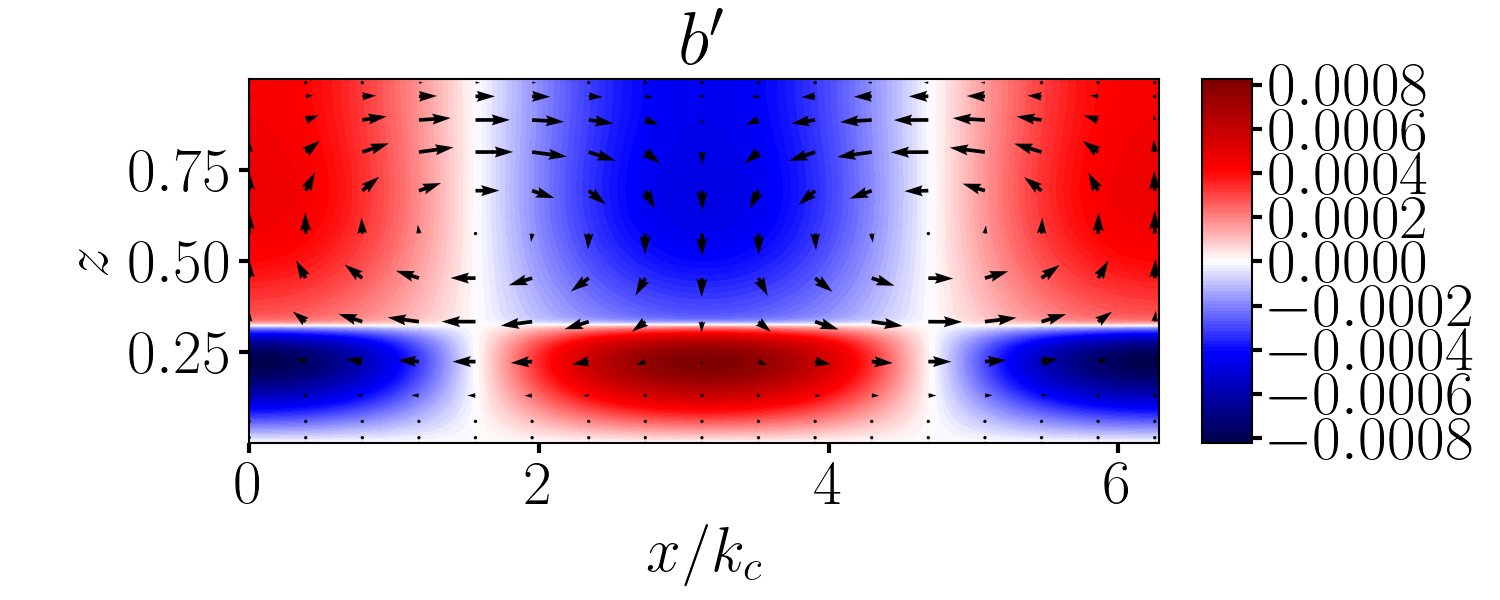}
  \\
  \includegraphics[width=0.69\textwidth]{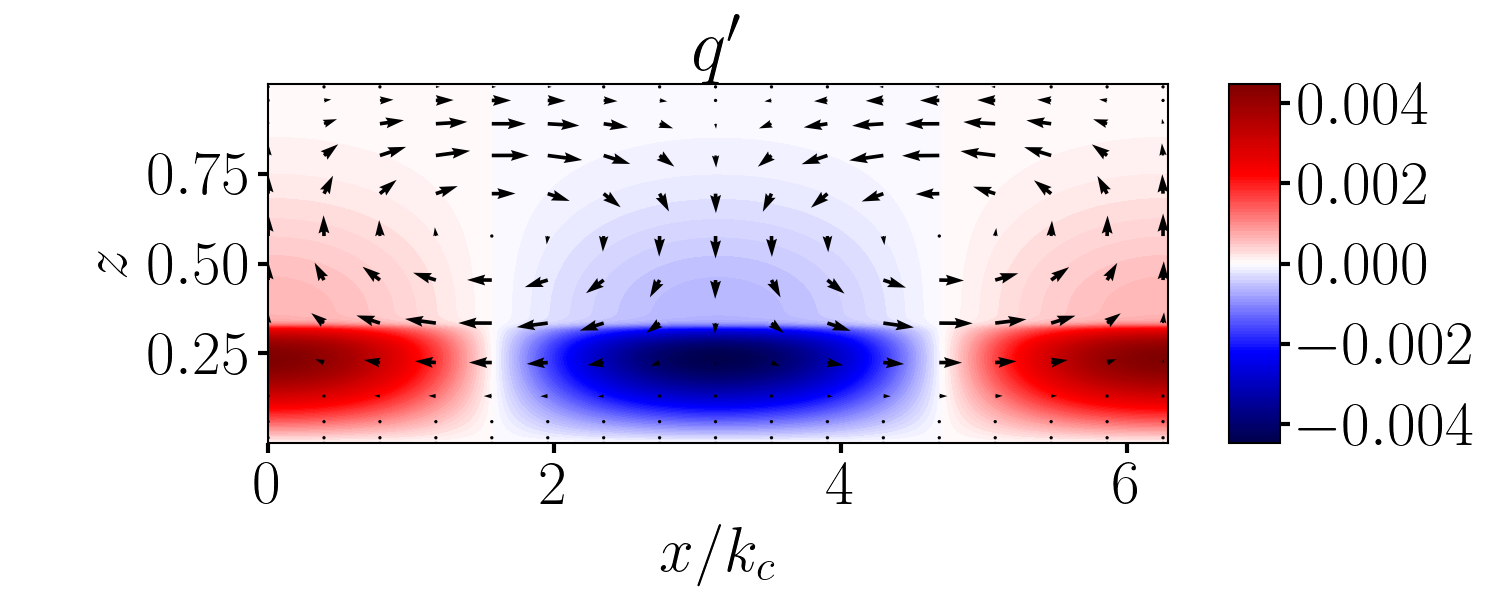}
  
  \caption{The real parts of the buoyancy and specific humidity perturbation quantities at criticality. The perturbation quantities are calculated using $f'(x,z) = \hat{f}(z) e^{ik_c x}$, where $\hat{f}(z)$ denotes the eigenvector of the quantity $f$. The velocity is shown by the black arrows. The results are shown for the parameter values: $RH_{\textsf{surf}} = 0.6$, $b_{\textsf{surf}} = 0.05$ and $r = 1\times10^{-5}$ and $Ra = Ra_c = 7.80\times10^5$.}
\label{fig:base_modal}
\end{center}
\end{figure}
The results in Figures \ref{fig:base_evs}, \ref{fig:base_modal} $\&$ \ref{fig:base_cond} indicate that the linear perturbation causes instability to occur in the updraft regions around the LCL, by consuming pCAPE in the upper regions ($z \geq z_s$), moistening the overall column and reducing the lower level inhibition by reducing the CIN and the LFC, and increasing pCAPE in a small vertical region between the LFC and the LCL. The moistening causes the largest increase in the condensation term around the LCL of the environment, indicating that the instability stems from this region of the domain.

\begin{figure}
\begin{center}
  \includegraphics[width=0.69\textwidth]{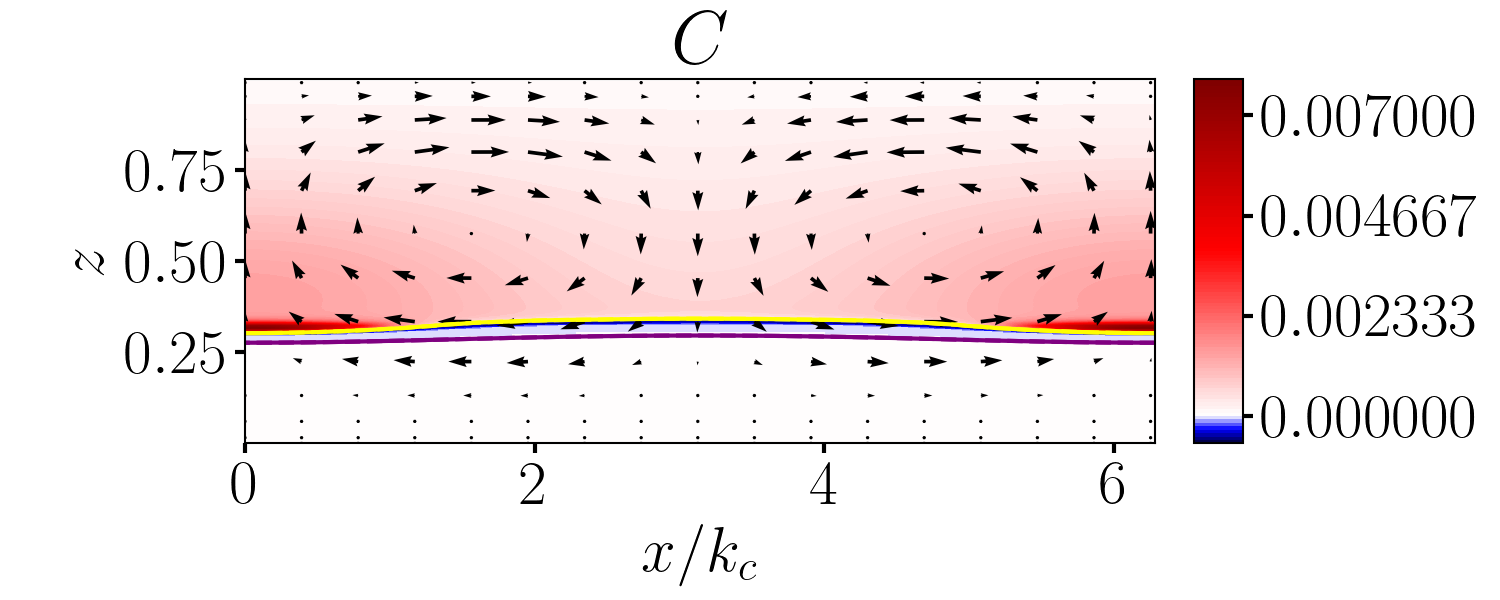}
  \\
  \includegraphics[width=0.69\textwidth]{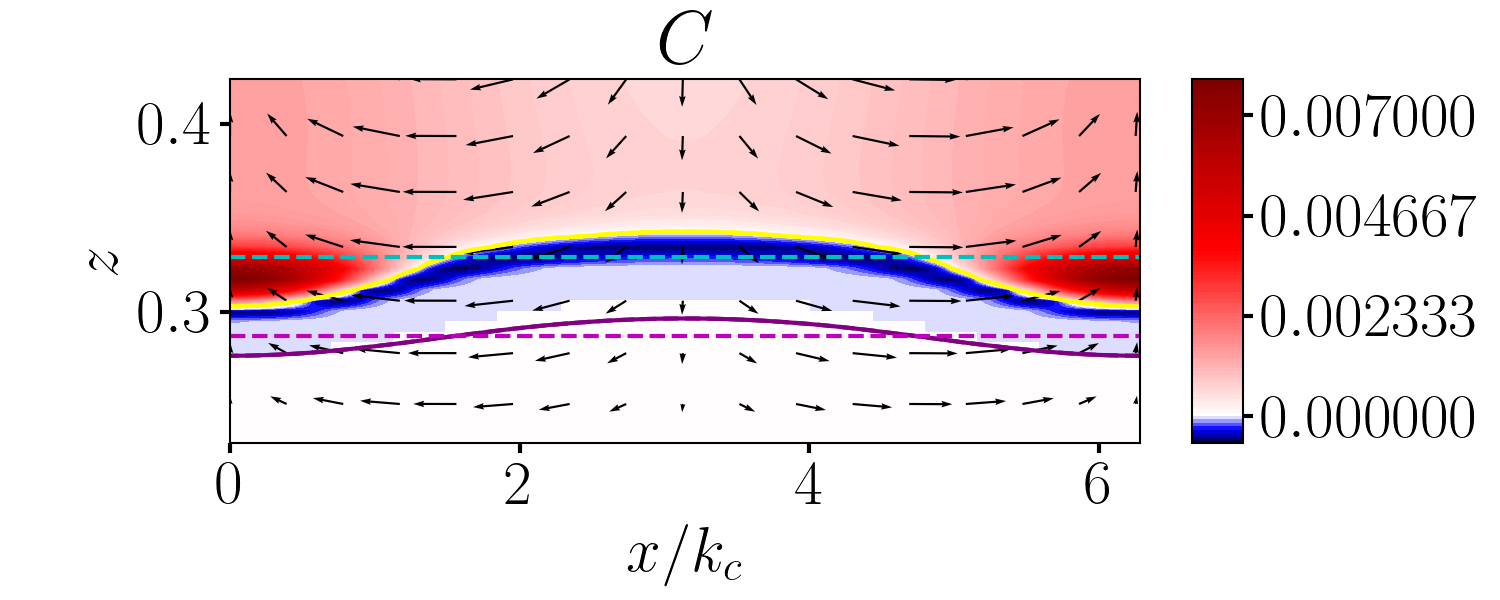}
  
  \caption{Condensation term $C = \gamma(q-q_s)\mathcal{H}(q-q_s)/\tau$, shown over one horizontal wavelength. The top panel shows $C$ for the full vertical domain, and the bottom panel shows a zoom of $C$ around the LCL of the environment, where the impact of the linear instability on the basic state is most pronounced. The cyan and magenta line mark the LCL and LFC of the environment respectively (with the dashed lines marking the basic state values). The grey arrows show the velocity $(u,w)$. The results are shown for the parameter values: $RH_{\textsf{surf}} = 0.6$, $b_{\textsf{surf}} = 0.05$ and $r = 1\times10^{-5}$ and $Ra = Ra_c = 7.80\times10^5$.}
\label{fig:base_cond}
\end{center}
\end{figure}

\subsection{Approximate Dispersion Relationship for Highest Frequency Modes}
\label{approx disp section}
The eigenvalue spectra shown in Figure \ref{fig:base_spectra} revealed the presence of damped oscillatory modes. \cite{oishi2024_lsa} assumed that the oscillatory modes are dry Boussinesq internal gravity waves, and examined the associated eigenvector structure of the modes to obtain an estimate for the moisture modified internal gravity waves. We also assume the waves are dry internal gravity waves, but use a different vertical wavenumber estimate motivated by the structure of the $b$ and $q$ eigenvectors.
\begin{figure}[h]
\begin{center}
  \includegraphics[width=0.495\textwidth]{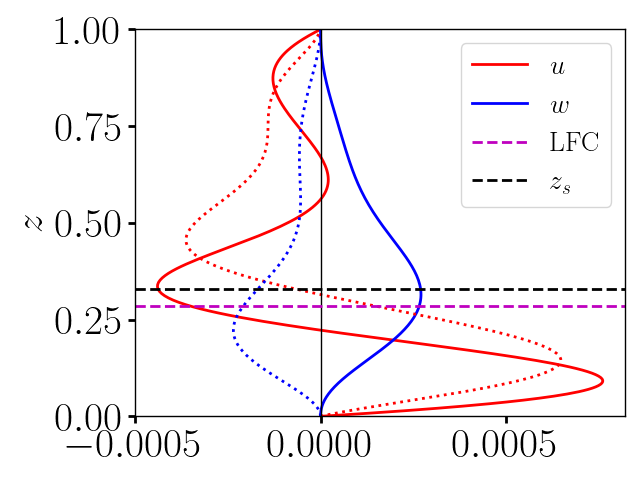}
  \includegraphics[width=0.495\textwidth]{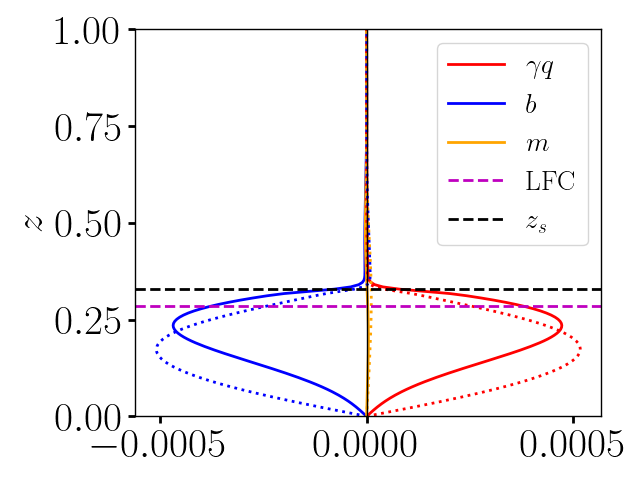}
  
  \caption{An example of the normalised eigenvectors at criticality, $Ra=Ra_c \;\&\; k=k_c$, associated with the highest frequency mode. Note that the solid and dotted lines represent the real and imaginary parts of the eigenvectors, respectively. The left panel shows the $u$ and $w$ eigenvectors, whilst the right panel shows the $b$, $\gamma q$ and $m$ eigenvectors. The dashed lines mark the basic state LCL ($z_s$, red) and the LFC (blue). The results are shown for the parameter values: $RH_{\textsf{surf}} = 0.6$, $b_{\textsf{surf}} = 0.05$ and $r = 1\times10^{-5}$ and $Ra = Ra_c = 7.80\times10^5$.}
\label{fig:wave_evs}
\end{center}
\end{figure}
\\
\newline
The highest frequency normalised eigenvectors shown in Figure \ref{fig:wave_evs}. The buoyancy and humidity eigenvectors are trapped below the LCL, in the unsaturated region of the domain. The imaginary part shown by the dotted lines corresponds to a tilt of the modal structure.
\\
\newline
Recall that the dispersion relationship for Boussinesq internal gravity waves is given by,
\begin{equation*}
    \omega_r = \frac{N_b k_x}{\sqrt{k_x^2 + k_z^2}},
\end{equation*}
 then we can get an approximate dispersion relationship for $\omega_r$ by estimating $N_b$ and $k_z$. We estimate the Brunt–Väisälä frequency by $N_b = \partial b / \partial z (z=0)$, where $b$ is the full buoyancy field. The structure of the $b \;\&\; q$ eigenvectors motivate choosing a vertical wavenumber of $k_z = \pi/z_s$, since there is half a wave under the LCL. Note that in \cite{oishi2024_lsa}, the vertical wavenumber was chosen to be $k_z = 2\pi/L_z$, where $L_z = 1$ is the domain height. We can write the approximate dispersion relationship for the moisture modified internal gravity waves as,
 \begin{equation}
     \omega_r \approx \frac{\partial b / \partial z \rvert_{z=0}  \:k_c}{\sqrt{k_c^2 + (\pi/z_s)^2}}
 \end{equation}
Note that in the non-linear system, internal gravity waves cause triggering of convection and are themselves generated by convective plumes \citep{vallis2019simple}. We examine the approximate relationship across the parameter space in the following section.

\subsection{Parameter Dependence of the Linear Perturbation}
\begin{figure}[h]
    \centering
    \includegraphics[width=0.99\textwidth]{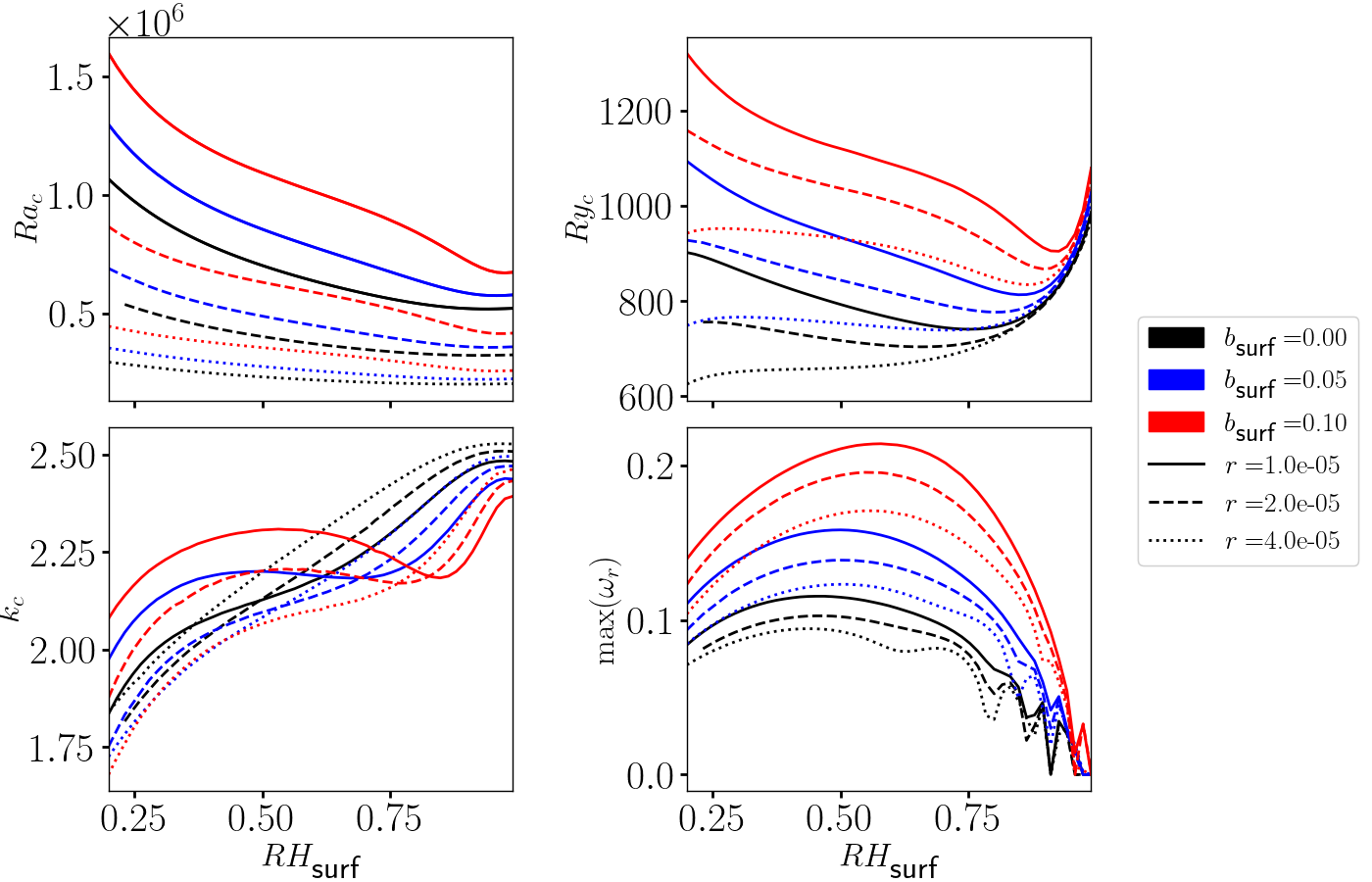}
    \caption{Critical Rayleigh (top left), Rainy (top right) and wavenumber (bottom left), and the highest frequency (bottom right) as functions of the surface relative humidity. The different line colours represent different surface temperature increases, and the different line styles represent different radiative cooling rates. The idealised climate change scenario involves going from the black solid line to the red dashed line, keeping the surface relative humidity fixed.}
    \label{fig:crit}
\end{figure}

To investigate how the values of the climate parameters change the action of the linear perturbation (at criticality) on the system, we examine changes in the critical parameters, the action on conditional instability, and in the structure of the buoyancy and moisture fluxes as parameters are varied.
\\
\newline
We first look at how the values of the critical Rayleigh number, Rainy number and wavenumber change with the parameters. Figure \ref{fig:crit} shows $Ra_c$, $Ry_c$, $k_c$ and $\max(\omega_r)$ as a function of the surface relative humidity, for various different surface temperature increases and radiative cooling rates. Recall that the value of $Ra$ only reflects the value of diffusion in our system; it cannot change according to changes in moist stability, and so it fundamentally fails to capture the moist convective behaviour. The critical Rainy number varies much less than the Rayleigh number across the climate parameter space, as a consequence of the incorporating a quantification of moist stability in its definition: $Ra_c$ varies by a factor of $\sim 8.0$ across all of the parameter values, whereas $Ry_c$ varies by a factor of $\sim 2.1$. The Rainy number captures the ratio of conditional instability (quantified by pCAPE and its associated length scale) to diffusion, and it therefore follows that the Rainy number is the more useful parameter for describing the state of the moist system. 
\\
\newline
For the idealised climate change scenario  (comparing the black solid line to the dashed red line, for fixed $RH_{\textsf{surf}}$) we see an increase in $Ry_c$ in Figure \ref{fig:crit}. The critical Rainy number increases as a result of increased moisture levels (associated with warmer surface temperatures): both the parcel and (basic state) environment profiles become warmer, increasing curvature of the moist pseudoadiabat, and result in an increase in CIN (and the LFC - see Figure \ref{fig:bprofs}). With increased levels of inhibition, pCAPE must be higher for the system to become unstable, and so $Ry_c$ increases under climate change. The increased levels of inhibition and the higher levels of available potential energy point to an intensified water cycle, characterised by stronger ($w^2/2 \sim$ pCAPE) more intermittent (higher CIN) convection. It is worth noting that $Ry_c$ is independent of $b_{\textsf{surf}}$ and $r$ in the saturated limit, however $Ra_c$ varies significantly, indicating that pCAPE $\sim 1/Ra_c$ as $RH_{\textsf{surf}} \rightarrow 1$ (since the LFC $\rightarrow 0$). 
\begin{figure}[h]
    \centering
    \includegraphics[width=0.99\textwidth]{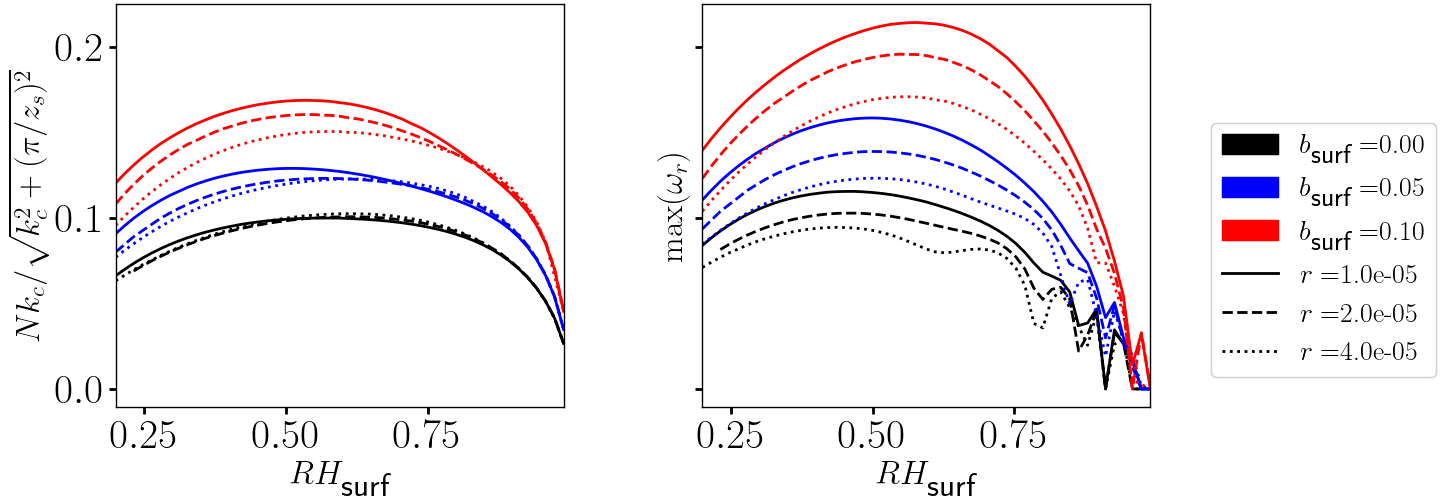}
    \caption{Dispersion relationship for the highest frequency mode (left), and the highest frequency mode calculated by solving the eigenvalue problem (right). The curves are all calculated at criticality ($Ra=Ra_c$). The different line colours represent different surface temperature increases, and the different line styles represent different radiative cooling rates. The idealised climate change scenario involves going from the black solid line to the red dashed line, keeping the surface relative humidity fixed.}
    \label{fig:crit_disp}
\end{figure}
\\
\newline
The critical wavenumber provides information about the width of the linear modes (i.e. updraft and subsiding regions). Note that the width of the updraft and subsiding regions are necessarily equal in the linear analysis, however this is not typically observed in non-linear simulations \citep{Agasthya_2024}. The linear theory does provide an initial indication of how we may expect convective plume widths to scale with the climate parameters. Examining the bottom left panel of Figure \ref{fig:crit}, we see that under the climate change scenario, if $RH_{\textsf{surf}} > 0.6$, $k_c$ decreases and the updraft and subsiding regions get wider. However, if $RH_{\textsf{surf}} < 0.6$, we see the opposite effect, with the updraft and subsiding regions becoming narrower. 
\\
\newline
Recall that the waves associated with the damped maximum frequency mode are damped (linear) dry internal gravity waves. The bottom right panel of Figure \ref{fig:crit} shows the (maximum) frequency associated with the dry internal gravity waves, as a function of the parameters. We see that under a typical climate change scenario (going from black solid to red dashed line, with $RH_{\textsf{surf}}$ fixed), the maximum frequency increases, which is associated with an increase in the triggering of convection. However, the triggering is reduced by increased CIN levels in the lower unsaturated region of the domain under climate change. In Section \ref{approx disp section}, we derived an estimate for the frequency of these moisture modified internal gravity waves. That approximate dispersion relationship is shown against the calculated maximum frequency values in Figure \ref{fig:crit_disp}. We see a qualitative agreement between the approximate and actual values of the maximum frequency. The approximate maximum frequency does tend towards zero in the saturated limit, and also captures the decrease in frequency as $RH_{\textsf{surf}} \rightarrow 0.25$. There are discrepancies in the position and magnitude of the peak between the approximate and actual values of the maximum frequency. Note that the dispersion relationship is the best fit for the lines with $b_{\textsf{surf}} = 0$, in which domain moisture is at a minimum (relative to the other curves). The moisture modified internal gravity waves are responsible for triggering convection in the non-linear system \citep{vallis2019simple}, and so understanding how their frequency changes under climate change gives us some initial insight into how we may expect the non-linear behaviour to respond.
\\
\newline
We could regard the (most unstable) linear mode as large scale on a global domain, and consider how it would effect the conditional instability of embedded storms. The effect of the linear perturbation on the conditional instability is shown by looking at the structure of the buoyancy eigenvectors. Since the parcel profile is not influenced by $\hat{b}$, the vertical regions where $\hat{b} < 0$ cool the environment and result in a reduction of inhibition (if $b_p < b_E$) or increase in instability (if $b_p > b_E$) in the region. Similarly, the vertical regions where $\hat{b} > 0$ cause the environment to warm, which causes an increase in inhibition or a decrease in instability in that region. The buoyancy eigenvectors shown in Figure \ref{fig:b_evs} represent the linear perturbation updraft regions. The action of the linear perturbation in an updraft region on the conditional instability remains the same, for each of the different parameter values: $\hat{b} < 0$ in the lower region, which extends to just below $z_s$ (and just above the LFC). There is a reduction in the pCAPE in the upper region where $\hat{b} > 0$, and a sharp change in the gradient of $\hat{b}$ (and $\hat{q}$) just above $z_s$. 
\begin{figure}[h]
\begin{center}
  \includegraphics[width=0.495\textwidth]{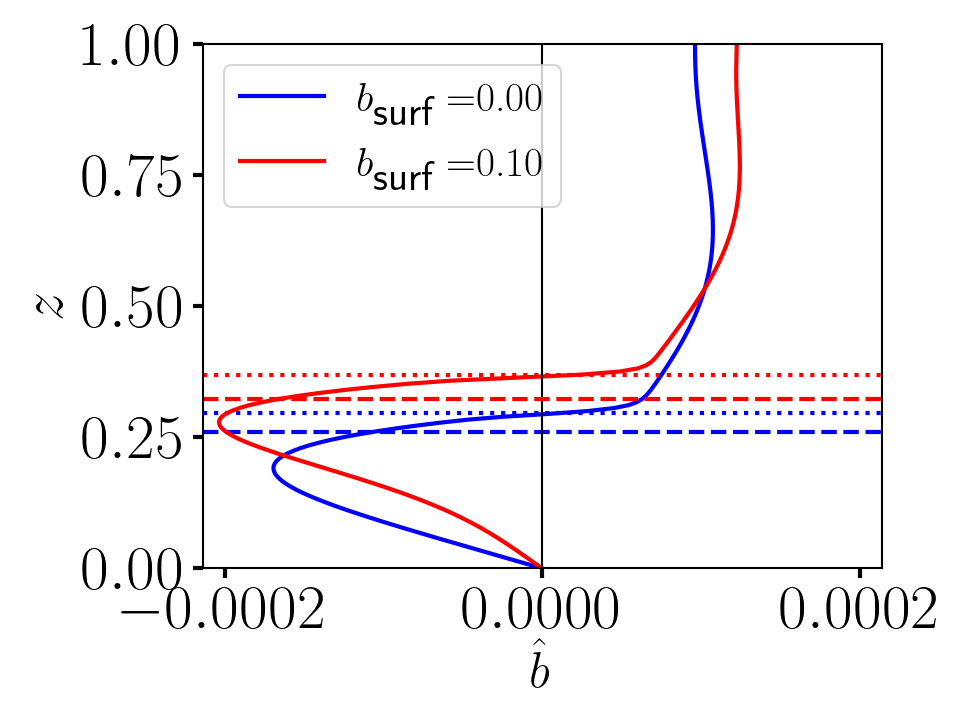}
  \includegraphics[width=0.495\textwidth]{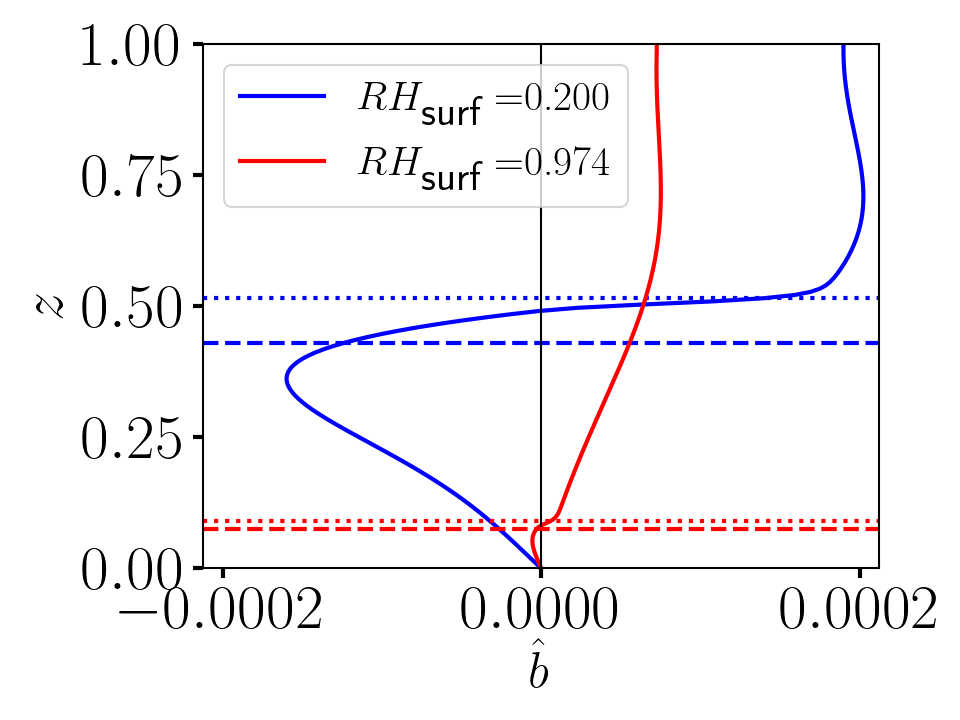}
  \\
  \includegraphics[width=0.495\textwidth]{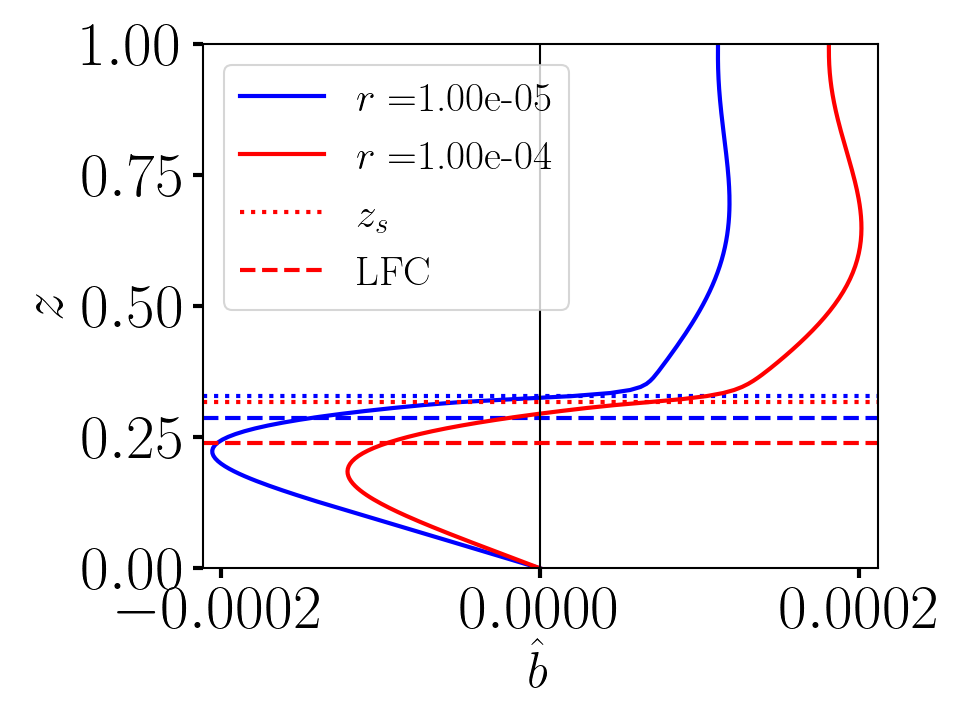}
  
  \caption{Buoyancy eigenvectors at criticality, plotted for varying: surface temperature (top left), surface relative humidity (top right), and radiative cooling (bottom). Dashed lines mark the LFCs and dotted lines mark the LCLs of the different environments. The eigenvectors are normalised such that $\max(w_r) = 5\times10^{-4}$ and $w_i = 0$.}
\label{fig:b_evs}
\end{center}
\end{figure}
\\
\newline
\begin{figure}[h]
    \centering
    \includegraphics[width=0.495\textwidth]{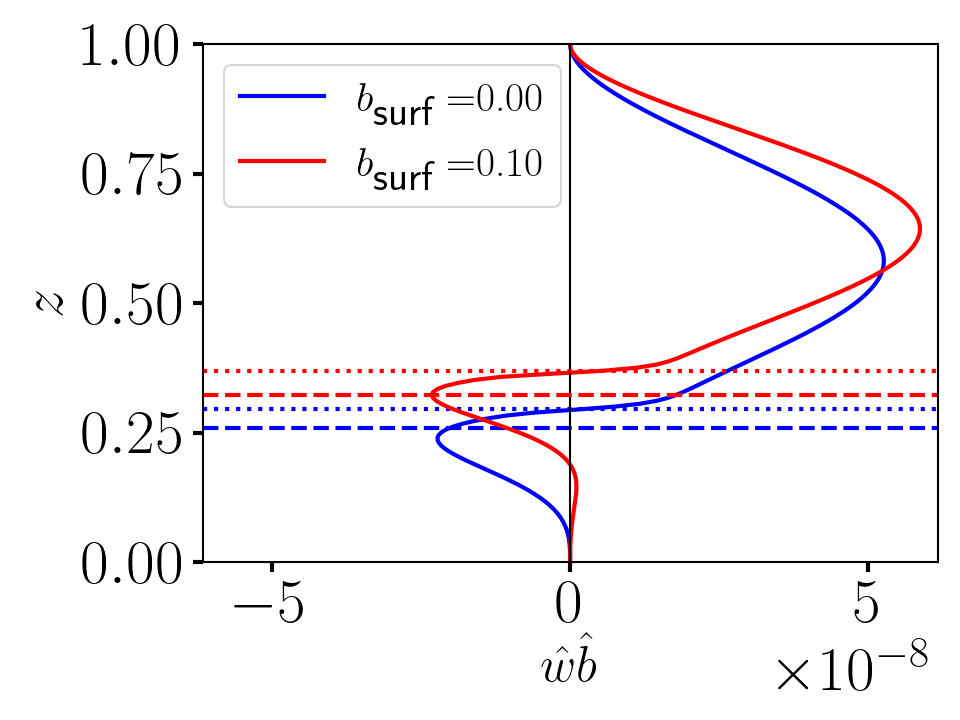}
  \includegraphics[width=0.495\textwidth]{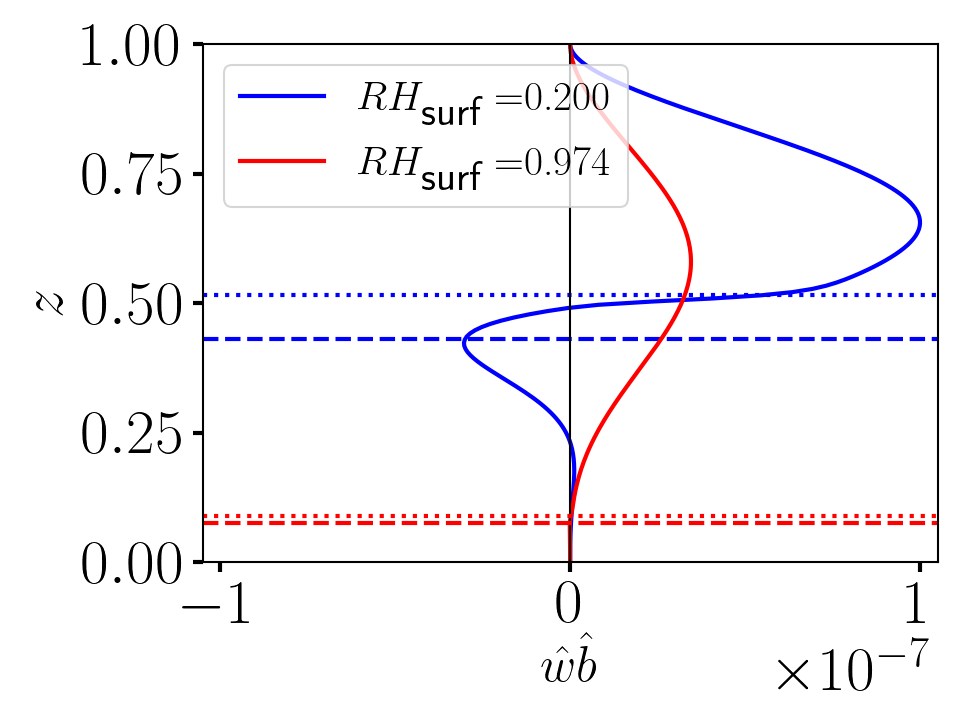}
  \\
  \includegraphics[width=0.495\textwidth]{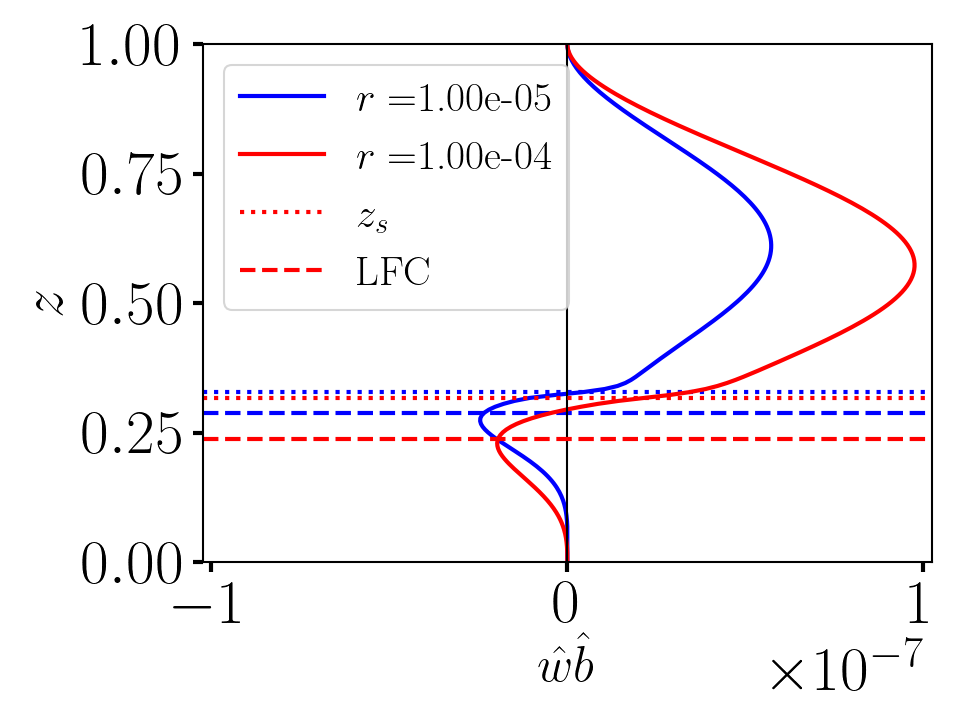}
  
  \caption{Buoyancy flux $w'b'$ at criticality, plotted for varying: surface temperature (top left), surface relative humidity (top right), and radiative cooling (bottom). Dashed lines mark the LFCs and dotted lines mark the LCLs of the different environments. The eigenvectors are normalised such that $\max(w_r) = 5\times10^{-4}$ and $w_i = 0$.}
    \label{fig:wb_flux}
\end{figure}
\begin{figure}[h]
    \centering
    \includegraphics[width=0.495\textwidth]{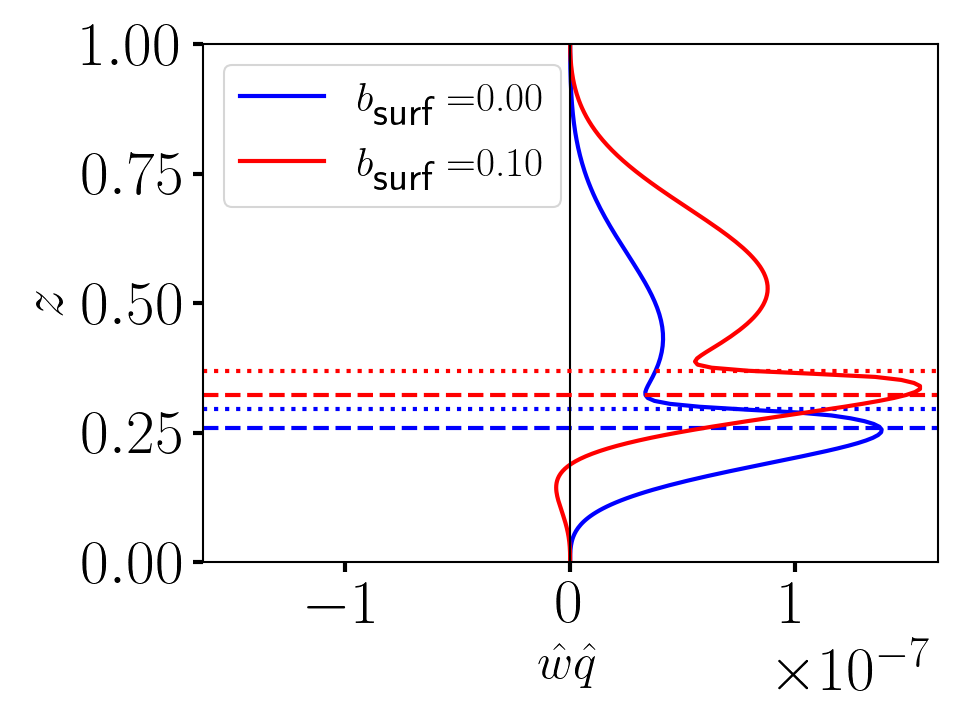}
  \includegraphics[width=0.495\textwidth]{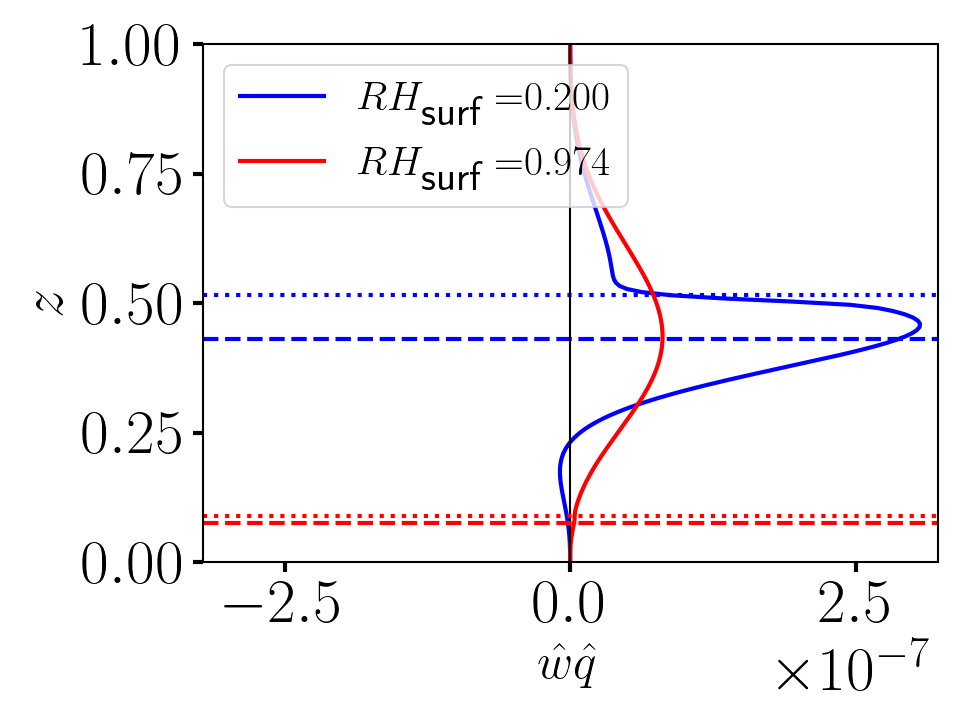}
  \\
  \includegraphics[width=0.495\textwidth]{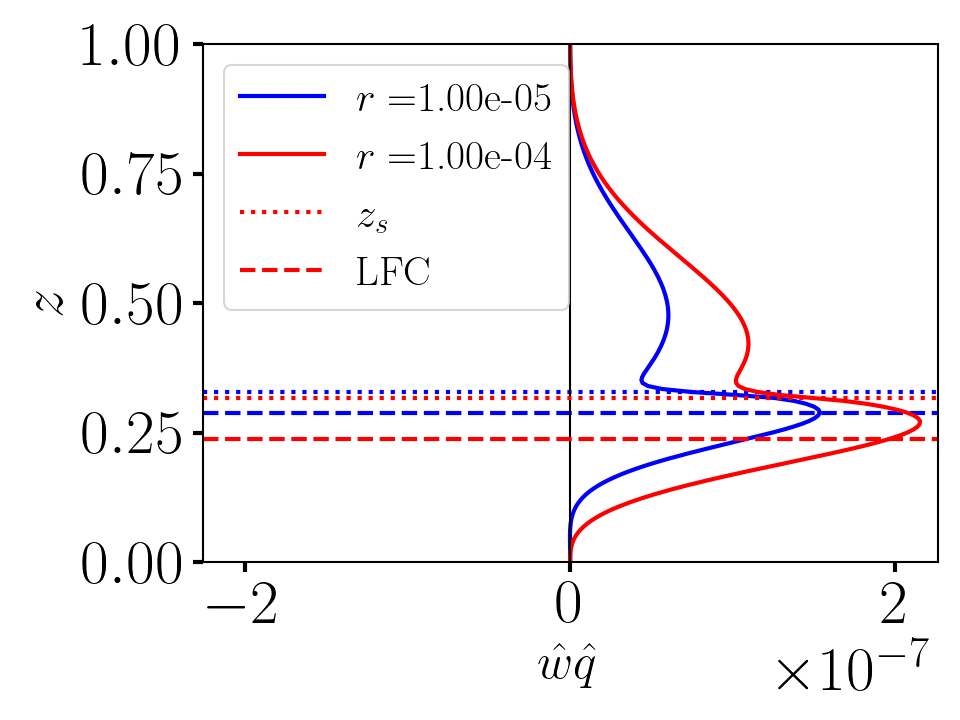}
  
  \caption{Specific humidity flux $w'q'$ at criticality, plotted for varying: surface temperature (top left), surface relative humidity (top right), and radiative cooling (bottom). Dashed lines mark the LFCs and dotted lines mark the LCLs of the different environments. The eigenvectors are normalised such that $\max(w_r) = 5\times10^{-4}$ and $w_i = 0$.}
    \label{fig:wq_flux}
\end{figure}
The buoyancy and moisture fluxes can be used to give an initial indication of the non-linear transport of buoyancy and moisture. Figure \ref{fig:wb_flux} shows how the horizontally averaged vertical transport of the buoyancy perturbation ($\hat{w}\hat{b} \sim \langle w' b' \rangle$) changes with the climate parameters. Apart from the $RH_{\textsf{surf}} = 0.974$ case, all of the buoyancy fluxes $w'b'$ display a similar structure: they have a negative peak of flux in the lower region, which occurs just below the LFC and a larger positive peak in the upper region. The height of the positive peak is dependent on $z_s$, such that if $z_s$ increases, we expect the height of the positive flux to increase with it. The nearly saturated $RH_{\textsf{surf}} = 0.974$ case does not exhibit a significant region of negative flux, which we conjecture is a result of the CIN being close to zero in this region. The perturbation moisture flux is shown in Figure \ref{fig:wq_flux}. Again, the nearly saturated case shows different behaviour than the rest of the parameter values displayed. The general behaviour of the moisture flux shows the peak moisture flux occurs around the LFC and below $z_s$, with a second smaller peak occurring in the upper saturated region. The $b_{\textsf{surf}} = 0.1 \; \& \; RH_{\textsf{surf}}$ cases show a small region close to the surface where the moisture flux towards the surface, which is caused by $w' < 0$ in this region due to the higher levels of CIN in these two cases. Note that the linear perturbation for the nearly saturated $RH_{\textsf{surf}} = 0.974$ case shows a different behaviour to the other cases, with the peaks in the moisture and buoyancy fluxes no longer occurring around the LFC, but around the middle of the domain. Examining the $Ry_c$ panel of Figure \ref{fig:crit}, we see that the critical Rainy number is almost independent of the surface temperature and radiative cooling rate for the nearly saturated regimes, which is a result of low levels of CIN and the LCL causing a different action of the linear perturbation.

\section{Discussion}
We have presented a detailed analysis of a simple framework for studying changes in moist convection under climate change. The Rainy-Be\'nard model is set up for climate forcing simulations, by adding a constant radiative cooling term to the buoyancy equation, and choosing appropriate boundary conditions: at the bottom boundary we impose the relative humidity and surface temperature, and at the top boundary we impose moist pseudoadiabatic boundary conditions which give the temperature, buoyancy and humidity at the top boundary freedom to adjust. We also impose idealised no-slip boundary conditions at both boundaries. Climate change can be imposed by varying the climate parameters, namely the surface temperature, the radiative cooling rate and the surface relative humidity: we illustrate a typical climate change scenario by doubling the radiative cooling in response to a $10K$ increase in surface temperature, keeping the surface relative humidity fixed.
\\
\newline
The fundamental linear behaviour of the Rainy-B\'enard model has been studied using a basic state analysis (Section 3) and a linear stability analysis (Section 4), and quantified using both moisture and conditional instability diagnostics. The basic state solution was found analytically to higher accuracy than in \cite{oishi2024_lsa}, so that the condensation and precipitation are non-zero (consistent with the budgets derived in Section 2.3). The pseudoadiabtic boundary conditions taken at the top boundary allow realistic adjustment in the basic state solution to close to the neutral parcel profile (Figure 4), which is a well-observed feature of tropical convective environments (e.g. \cite{Betts_1986}) and was not possible in the previous studies by \cite{vallis2019simple}, \cite{Agasthya_2024}, and, \cite{oishi2024_lsa}. The basic state analysis reveals that the radiative cooling parameter is primarily responsible for changes in the (basic state) conditional instability, whereas the surface temperature and surface relative humidity are responsible for changes in the (basic state) precipitation (Figures 5 and 6).
\\
\newline
We used a linear instability analysis to calculate the critical Rayleigh number for convective onset for a range of climate change parameter values, examining the action of the most unstable mode (Section 4.1) and the waves associated with the highest frequency mode. The most unstable mode has a moistening action on the updraft ($w' > 0$) regions of the domain, and a drying action on the subsiding regions ($w' < 0$) of the domain, with a circulation from the moister updraft regions to the drier subsiding regions (Figures 8, 9 and 10). In terms of conditional instability, we find that the most unstable mode causes a reduction of both the convective inhibition (CIN) below the LFC, and the pCAPE above the LFC, in the updraft regions. We find damped oscillatory modes associated with dry internal gravity waves trapped in the lower unsaturated region of the domain, as in Oishi and Brown 2025. We derived an approximate dispersion relationship for the dry internal gravity waves in equation (43), which differs from that given in Oishi and Brown 2025, but shows qualitative agreement with the numerically calculated results (Figure 13).
\\
\newline
A key result is the derivation of non-dimensional parameters which correctly capture the relationship between moist instability and diffusion (Section \ref{CAPE non-dim} and Appendix \ref{radiative Ry}). We used the positive convective available potential energy (pCAPE) as a scale for the kinetic energy in the system, and its associated length scale (the difference between the level of neutral buoyancy (LNB) and the level of free convection (LFC)) to construct a moist Rayleigh number, called the Rainy number:
\begin{equation*}
  Ry = \frac{\textrm{pCAPE} \times (\text{LNB} - \text{LFC})^2}{\nu \kappa}
\end{equation*}
The Rainy number represents the ratio of (moist) conditional instability (quantified by pCAPE) to diffusion. We find that $Ry$ changes in response to changes in moist instability; $Ra$ does not (Figures 1 and 12). By using conditional instability (quantified by pCAPE) to set the scales in the system, the Rainy number is a better control parameter for our model: across the climate parameter space, $Ry$ varies less at criticality than $Ra$. 
\\
\newline
Due to the relationship between radiative cooling and pCAPE in our solutions, we are also able to define a radiatively-based Rainy number (see Appendix \ref{radiative Ry}). The Radiative Rainy number, which uses a different measure of moist conditional instability, also varies less than $Ra$ at criticality as a result. In fact the use of the CAPE-based or radiatively-based Rainy number is probably an open choice according to the scientific question at hand. CAPE-based $Ry$ would naturally be appropriate for studies of convection and the water cycle; radiatively-based $Ry$ would be a natural
control for the system when studying sensitivity to boundary fluxes and radiative forcing.
\\
\newline
Under the typical climate change scenario, the basic state analysis results indicate that we expect more intense moist convection, with more rainfall. There is more moisture in the system (associated with warmer surface temperatures) and this leads to a mid-domain profile which is warmer, following a profile which is more ``bowed" (Figure \ref{fig:bprofs}), so despite the warmer and humid surface air, the warmer mid-levels lead to increased CIN under climate change. For the onset of convection to occur in a system with more CIN, the available potential energy (pCAPE) must increase, and as a result the critical Rainy number increases. The increase in the critical Rainy number is associated with an intensification of the water cycle (as found in \cite{CP4}): we expect more intense (increased pCAPE $\sim w^2/2$), more intermittent (increased CIN) moist convection under climate change. Our future research will build on this framework and examine non-linear simulations and transport of moist Rainy-B\'enard convection under climate change.

\subsubsection*{Acknowledgements}
The authors would like to thank Kasia Nowakowska, Geoff Vallis, Jeff Oishi, Ben Brown, Lokahith Agasthya and David Dritschel for helpful discussions on Rainy-B\'enard convection. This work was supported by the Leeds-York-Hull Natural Environment Research Council (NERC) Doctoral Training Partnership (DTP) Panorama under grant NE/S007458/, UK Met Office CASE project funding, and by the LMCS project under NERC grant, NE/W001888/1. This work was undertaken on ARC4, part of the High Performance Computing facilities at the University of Leeds, UK.

\newpage
\appendix
\section{CAPE for the Dry System}
From Wallefe and Smith 2015, the classic Rayleigh-B\'enard convection equations are:
\begin{align}
  \frac{D\bu}{Dt} &= - \grad \phi + g \alpha_v T \mathbf{k} + \nu \lap \bu, \\
  \frac{DT}{Dt} &= \kappa \lap T.
\end{align}
Note that temperature here is equivalent to buoyancy in our model. A surface parcel would maintain its temperature on adiabatic ascent ($DT/Dt = 0$), and so the parcel profile is given by $T_p(z) = T_0$. The basic state solution, for fixed temperature boundary conditions $T(0) = T_0, \: T(H) = T_1$ is,
\begin{equation}
  T = T_0 - \frac{\Delta T}{H}\!z,
\end{equation}
where $\Delta T \equiv T_0 - T_1$. As in Section 2.4, we take the environment to be the basic state, and we calculate the (dry) CAPE as:
\begin{equation}
  \textrm{CAPE} = \int^{H}_{0} (T_p - T_E) dz = \frac{\Delta T}{H} \int^{H}_{0} z \: dz = \frac{H \Delta T}{2}
\end{equation}
Note that $T_p \geq T_E$ at all heights, so $\textrm{pCAPE} = \textrm{CAPE}, \: \textrm{CIN} = 0, \: \textrm{LFC} = 0, \:\textrm{and}\: \textrm{LNB} = H$. Recalling the expression for the Rainy number (Equation (26)), it follows that,
\begin{equation}
  Ry = \frac{H^3 \Delta T}{2\kappa \nu} = \frac{Ra}{2 g \alpha_v}.
\end{equation}
Therefore, the Rainy number is directly proportional to the classical Rayleigh number for the dry system.

\section{Radiative Rainy Number}
\label{radiative Ry}
We rescale the dry adiabatic non-dimensionalisation, using the minimum moist static energy gradient of the basic state to set the timescale, and the height of the domain to set the length scale. Note that, from equation (31),
\begin{equation}
  \frac{dm}{dz} = r \! Ra^{1/2}(z-1).
\end{equation}
Next, we recall the alternative definition of conditional instability is $dm/dz < 0$ and $db/dz > 0$, and note that $dm/dz$ is a minimum at $z=0$. Writing,
\begin{equation*}
  -\min\Bigg(\frac{dm}{dz}\Bigg) \sim \frac{[B]}{[L]} = \frac{1}{[t]^2},
\end{equation*}
and taking $[L] = 1$ (non-dimensional height of the domain), the new scales can be written as:
\begin{equation}
  [t] = \frac{1}{r^{1/2} \! Ra^{1/4}}, \quad [L] = 1, \quad [B] = r \! Ra^{1/2}, \quad [U] = \frac{[L]}{[t]} = r^{1/2} \! Ra^{1/4}.
\end {equation}
Note that the (instability) timescale decreases with $r$ and $Ra$, and the velocity and buoyancy scales increase with $r$ and $Ra$. After some algebra, the rescaled momentum equation can be written as:
\begin{equation}
  \frac{D\hat{\bu}}{D\hat{t}} = - \grad \hat{\phi} + \hat{b} \mathbf{k} + \frac{1}{Ry_R^{3/4}}\lap \hat{\bu},
\end{equation}
where the Radiative Rainy number is defined as,
\begin{equation}
  Ry_R \equiv r^{2/3} \! Ra.
\end{equation}

\begin{figure}[h]
\begin{center}
  \includegraphics[width=0.48\textwidth]{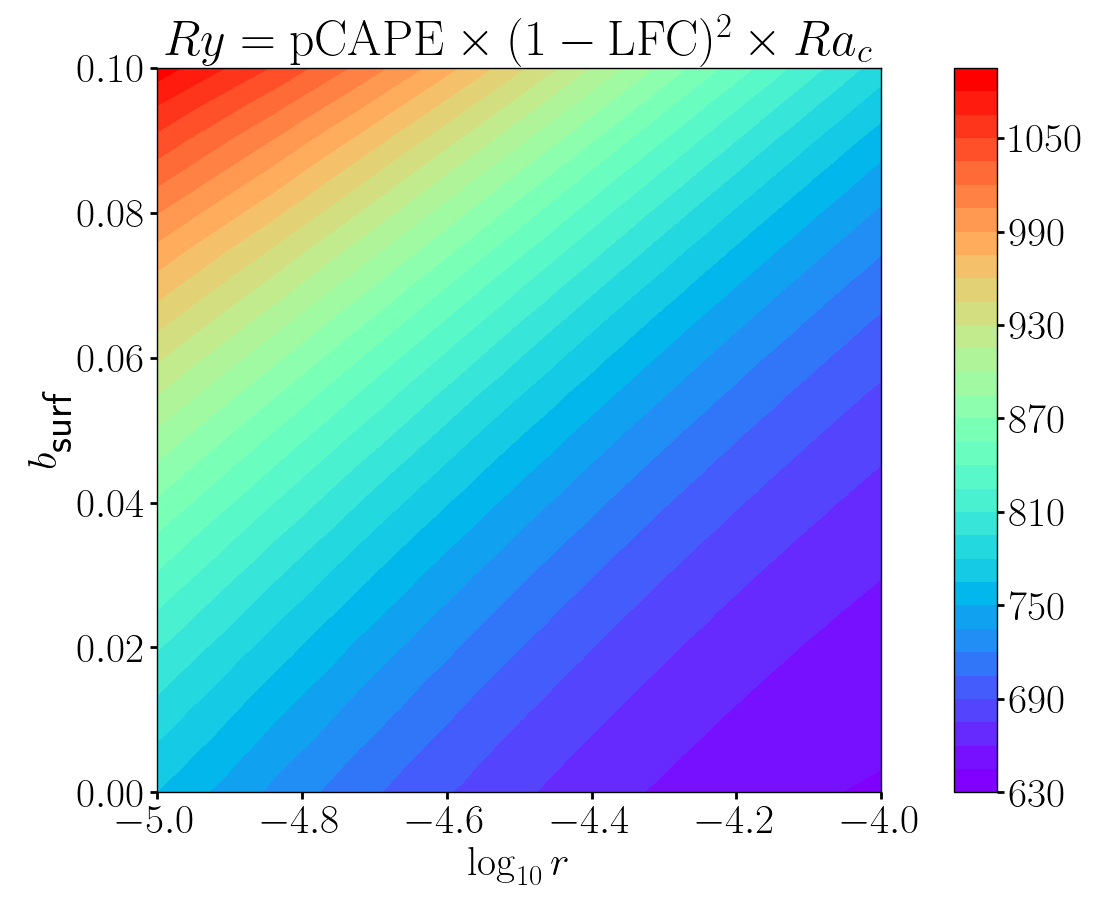}
  \includegraphics[width=0.48\textwidth]{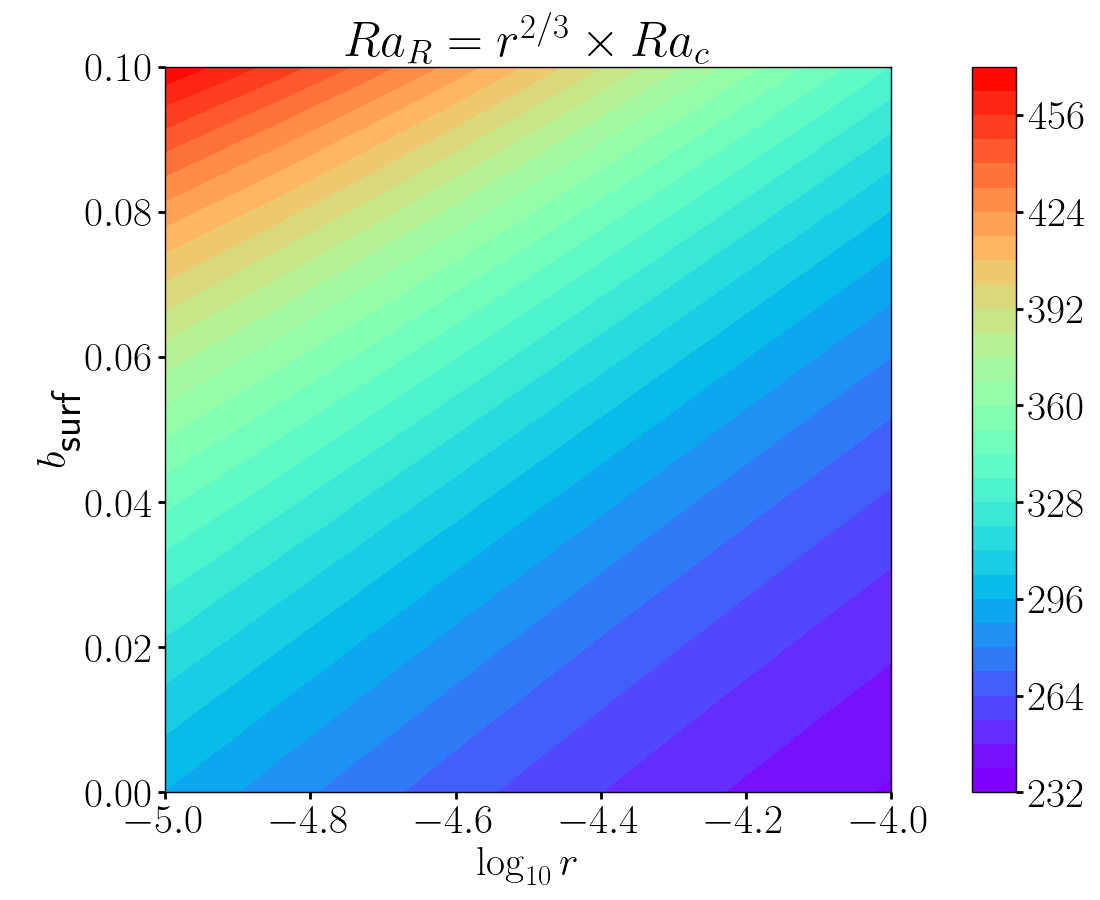}
  
  \caption{Critical Rainy number (left) and critical Radiative Rainy number (right). Both Rainy numbers are based on different quantifications of conditional instability, and show a degree of proportionality.}
\label{fig:Ry_vs_RaR}
\end{center}
\end{figure}
\begin{figure}[h]
\begin{center}
  \includegraphics[width=0.98\textwidth]{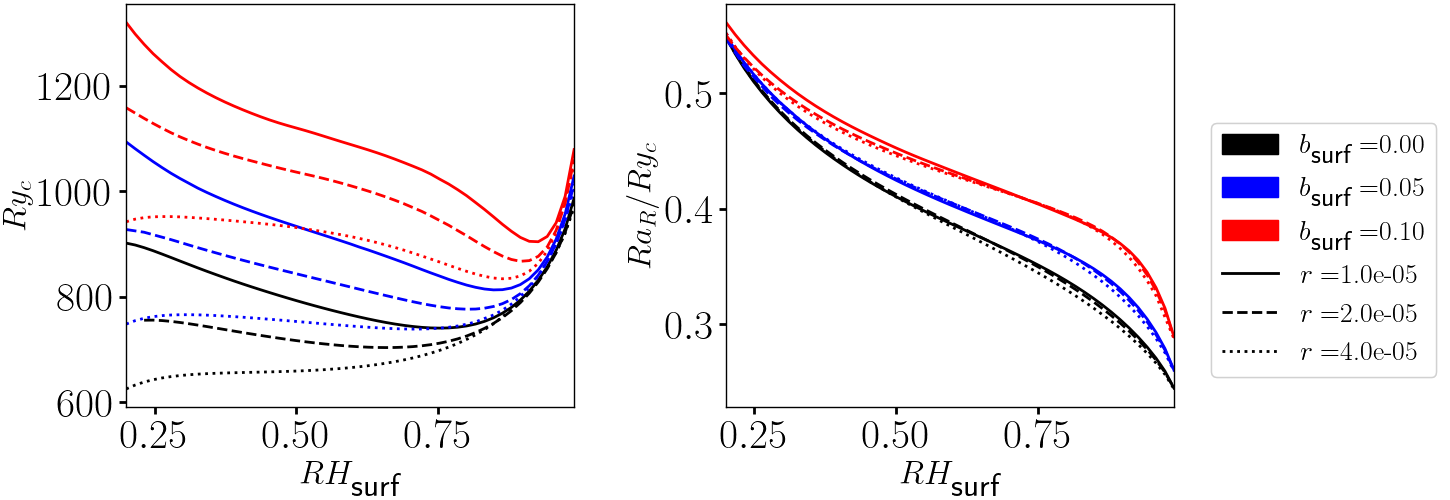}
  \caption{Ratio of the critical Rainy number to the critical Radiative Rainy number}
\label{fig:crit_Ry_vs_RaR}
\end{center}
\end{figure}

Note that $Ra_R/Ry$ at criticality shows a variation of $\sim 20\%$. This (approximate) proportionality reveals the relationship,
\begin{equation}
  \textrm{pCAPE} \sim \frac{r^{2/3}}{(1-\textrm{LFC})^2},
\end{equation}
at least for constant surface relative humidity ($RH_{\textsf{surf}} = 0.6$). The relationship for the different climate chance scenarios is shown in Figure \ref{fig:crit_Ry_vs_RaR}. The ratio does not vary significantly with radiation, relative to the changes with surface humidity or surface temperature.

\section{Parameter Correction}
\label{corected parameters}
In \cite{vallis2019simple}, the value of the constant $e_0 = 611 Pa$ in equation $(2.16)$ is valid for a reference temperature of $T_0 = 273 K$ \citep{CC_relationship}, rather than the specified reference temperature of $T_0 = 300K$ used throughout the rest of the paper. Letting $\theta_0 = 300 K$, $T_0 = 273 K$, and expressing $T = \theta_0 + \delta T$, we can write the saturation vapor pressure equation $(2.16)$ as,
\begin{equation}
    e_s = e_0 \exp \Bigg ( \frac{L}{R_v} \bigg\{ \frac{1}{T_0} - \frac{1}{\theta_0 + \delta T} \bigg\} \Bigg ).
    \label{sat vap pressure}
\end{equation}
Assuming $\theta_0 \gg \delta T$, equation \eqref{sat vap pressure} can be approximately written as,
\begin{equation}
    e_s \approx e_0 \exp \Bigg (\frac{L}{R_v} \bigg \{ \frac{\theta_0 - T_0}{T_0 \theta_0} \bigg \} \Bigg)\exp \Bigg ( \frac{L}{R_v} \bigg\{ \frac{\delta T}{T_0 \theta_0} \bigg\} \Bigg ).
    \label{lin vap pressure}
\end{equation}
Recall the approximate relationship between the saturation vapor pressure and saturation specific humidity given by equation $(2.18)$,
\begin{equation}
    q_s \approx \epsilon \frac{e_s}{p}.
    \label{sat q p relat}
\end{equation}
Using equation $(2.21)$, the pressure can be approximately written by,
\begin{equation}
    p \approx p_0 \exp \bigg ( \frac{c_p \delta T}{R_d \theta_0} \bigg ).
    \label{p eqn}
\end{equation}
Combining equations \eqref{lin vap pressure}-\eqref{p eqn}, the saturation vapor pressure can be expressed as,
\begin{equation*}
    q_s = \epsilon \frac{e_0}{p_0} \exp \Bigg (\frac{L}{R_v} \bigg \{ \frac{\theta_0 - T_0}{T_0 \theta_0} \bigg \} \Bigg)\exp \Bigg (\bigg\{\frac{L}{R_v \theta_0 T_0} - \frac{c_p}{R_d \theta_0} \bigg\} \delta T \Bigg ),
\end{equation*}
or,
\begin{equation}
    q_s = q_0 \exp{\big (\alpha \delta T \big)},
\end{equation}
where,
\begin{equation}
    q_0 = \epsilon \frac{e_0}{p_0} \exp \Bigg (\frac{L}{R_v} \bigg \{ \frac{\theta_0 - T_0}{T_0 \theta_0} \bigg \} \Bigg) = 0.019\: kg \, kg^{-1},
    \label{cor q0}
\end{equation}
and,
\begin{equation}
    \alpha = \frac{L}{R_v \theta_0 T_0} - \frac{c_p}{R_d \theta_0} = 0.054 K^{-1}.
\end{equation}
Note that in \cite{vallis2019simple}, $\alpha = 0.060 K^{-1}$, and $q_0 = 3.8 \times 10^{-3} \: kg \, kg^{-1}$, which is five times smaller than the value found in equation \eqref{cor q0}. The discrepancy in these parameters affects the values of the non-dimensional parameters $\alpha$ and $\gamma$ in the Rainy-B\'enard model.

\newpage
\noindent\bibliographystyle{gGAF}
\newline\bibliography{gGAFguide}
\vspace{12pt}

\end{document}